\documentclass[aps,pra,twocolumn,preprintnumbers,nofootinbib,amsmath,amssymb,superscriptaddress,showkeys,floatfix]{revtex4-2}

\usepackage[english]{babel}
\usepackage{amsmath}
\usepackage{amsfonts}
\usepackage{amssymb}
\usepackage{subfiles}
\usepackage{natbib}
\usepackage{mathtools}
\usepackage[colorlinks,linkcolor=red,citecolor=blue,breaklinks=true]{hyperref}
\usepackage{graphicx}
\usepackage{dcolumn}
\usepackage{bm}
\usepackage{color}
\usepackage{booktabs}
\usepackage{amsthm}
\usepackage{dsfont}
\usepackage{tcolorbox}
\usepackage{cases}
\usepackage{physics}
\usepackage{subfiles}

\usepackage{braket}
\usepackage{bbold}
\usepackage{xcolor}
\usepackage[normalem]{ulem} 
\usepackage{txfonts,times}
\usepackage{orcidlink}
\usepackage{lipsum}
\usepackage{tabularx}
\usepackage{booktabs}

\begin{document}
\date{December 25, 2025}

\title{Fluctuation theorems with optical tweezers: theory and practice}

\author{Thalyta T. Martins\orcidlink{0000-0003-2113-5468}}
\affiliation{Instituto de Física de São Carlos, Universidade de São Paulo, IFSC-USP, 13566-590 São Carlos, SP, Brasil}

\author{André H. A. Malavazi\orcidlink{0000-0002-0280-0621}}
\affiliation{%
International Centre for Theory of Quantum Technologies, University of Gdańsk, Jana Bażyńskiego 1A, 80-309 Gdańsk, Poland
}%

\author{Lucas P. Kamizaki\orcidlink{0000-0002-6485-713X}}%
\affiliation{Instituto de Física de São Carlos, Universidade de São Paulo, IFSC-USP, 13566-590 São Carlos, SP, Brasil}
\affiliation{Instituto de Física ‘Gleb Wataghin’, Universidade Estadual de Campinas, 13083-859 Campinas, SP, Brasil}

\author{Artyom Petrosyan}%
\affiliation{Université de Lyon, ENS de Lyon, CNRS, Laboratoire de Physique, F-69342 Lyon, France}

\author{Benjamin Besga}%
\affiliation{Université Claude Bernard Lyon 1, CNRS, institut Lumière Matière, UMR 5306, F69100 Villeurbanne, France}

\author{Sergio Ciliberto\orcidlink{0000-0002-4366-6094}}%
\affiliation{Université de Lyon, ENS de Lyon, CNRS, Laboratoire de Physique, F-69342 Lyon, France}

\author{Sérgio R. Muniz\orcidlink{0000-0002-8753-4659}}%
\affiliation{Instituto de Física de São Carlos, Universidade de São Paulo, IFSC-USP, 13566-590 São Carlos, SP, Brasil}

\begin{abstract}

Fluctuation theorems, such as the Jarzynski equality and the Crooks relation, are effective tools connecting non-equilibrium work statistics and equilibrium free energy differences. However, detailed hands-on, reproducible protocols for implementing and analyzing these relations in real experiments remain scarce. This tutorial provides an end-to-end workflow for measuring, validating, and applying fluctuation theorems using a single-beam optical tweezers setup. It introduces the foundational ideas and consolidates practical calibration (PSD-based trap stiffness and position sensitivity), protocol design (forward/reverse finite-time drives over multiple amplitudes and durations), and robust estimators for free-energy difference and dissipated work, highlighting finite-sampling and rare-event effects. We demonstrate the procedures using an extensive set of measured trajectories under different conditions and provide openly accessible datasets and Python code, enabling new researchers or educators to reproduce the results with minimal effort. Beyond pedagogical validation, we discuss how these recipes translate to broader soft‑matter and mesoscopic contexts. By combining user-friendly instruments with clear and transparent analysis, this work promotes the education and reliable adoption of stochastic thermodynamic methods in the curricula of physics and chemistry, as well as among emerging research teams.

\end{abstract}

\keywords{Jarzynski equality; fluctuation theorems; stochastic thermodynamics; optical tweezers}

\maketitle

\section{Introduction}

Thermodynamics is one of the most important and well-established physical theories \cite{zemansky, Callen, Prigogine}, with an impressive range of applications, especially in chemistry and engineering. Despite the initial pragmatic approach, with its humble beginnings in the phenomenological study of steam engines, and all the revolutionary developments in quantum physics, this classical theory still remains indispensable for describing energy exchange and dissipation. Its prominent position is due to the effective universality of its core principles. In this sense, the laws of thermodynamics have proved to be a solid basis for understanding nature, with applications ranging from chemistry and biology \cite{Onuchic1997, Qian2007, Garcia2011, England2013, England2016, RevModPhys.91.045004, Suma2023, Cao2025} to computer science and computation \cite{Parrondo2015,Wimsatt2021,Wolpert2024}, and even black-hole physics \cite{Bekenstein1972, PhysRevD.15.2738, Witten2025}.

The early discipline of classical thermodynamics, however, was restricted to describing macroscopic \textit{systems in equilibrium} and \textit{quasistatic processes}. Thermodynamic equilibrium refers to static (time-invariant) states of macroscopic variables such as volume, pressure, and temperature. Quasistatic processes are idealized transformations in which the system is kept in equilibrium throughout its evolution. Another central concept is the notion of reversibility, which assures that the system can be returned to a previous state by reverting some relevant parameters. This is intrinsically related to the condition of quasistatic transformations. For these conditions to hold, quasistatic processes must proceed very slowly because quick shifts in the system's configuration could push it out of equilibrium. Nevertheless, in truly macroscopic systems, equilibration usually occurs quickly enough to treat finite-time changes as approximately quasistatic. Notice that despite the term \textit{thermodynamics}, the foundational equilibrium theory does not portray dynamic processes in the usual sense. The dynamics is given through relations between macroscopic quantities, but the underlying evolution remains hidden. Thus, while effectively predicting the final state of a system, the theory lacks the tools to describe \textit{how}, \textit{at what rate}, or \textit{through which mechanisms} a system transitions to equilibrium. Capturing these dynamic aspects usually requires other frameworks to describe the connection to the underlying microscopic processes.

The advent of statistical physics began to bridge the gap with the microscopic nature of matter. This progress introduced probabilistic and statistical reasoning \cite{Reif-book,Huang-book,Salinas-book,Prigogine-book}, which proved highly successful in describing microscale physics and justifying previous phenomenological observations. Good examples are Einstein's explanation of Brownian motion and Perrin's confirmation of the atomic hypothesis \cite{einstein1905motion, perrin1909mouvement}. 
These developments opened the way for extensions to smaller systems far from equilibrium and finite-time protocols \cite{ritort2008nonequilibrium, Jarzynski2011, Seifert2012, ciliberto2017experiments, Seifert2019}. Under these conditions, fluctuations are non-negligible and play a significant role in the physical description. 
These ideas are now contained within the broader context of \textit{stochastic thermodynamics} 
\cite{Peliti-book,Seifert-book}, which provides a general framework to define quantities such as work, heat, and entropy, for mesoscopic systems following trajectories in phase space. In this context, these quantities are dynamic stochastic variables that can acquire different values for distinct process realizations, and familiar laws of thermodynamics are recovered at the average level over an \textit{ensemble} of realizations.

The fluctuation theorems (FTs) encapsulate some of the most important results of the field \cite{evans1993probability, evans2002fluctuation, gallavotti1995dynamical, lebowitz1999gallavotti, kurchan1998fluctuation, Jarzynski2006, searles2007steady, evans2008fluctuation, seifert2005entropy, puglisi2006relevance, Harris2007}. Among them, the most well-known were proposed by Jarzynski \cite{Jarzynski1997prl} and Crooks \cite{crooks1998nonequilibrium,crooks1999entropy} regarding non-equilibrium work relations. In essence, FTs offer general mathematical statements for the probability distribution of relevant thermodynamic quantities. It is fascinating how these simple relations carry profound conceptual meanings and far-reaching consequences despite their straightforward structures. Experimentally, they are well established and verified \cite{liphardt2002equilibrium, Wang2002, Carberry2004, collin2005verification, Blickle2006} on numerous platforms, such as electrical circuits \cite{vanZon2004, garnier2005nonequilibrium, Andrieux2007}, mechanical torsion pendulums \cite{douarche2005estimate}, and single-electron boxes \cite{koski2013distribution} (see, for instance, \cite{ciliberto2017experiments}). Furthermore, these results have been extended to the quantum regime \cite{tasaki2000jarzynski, kurchan2001quantum, PhysRevLett.92.230602, Campisi2009, Esposito2009, Campisi2011, funo2018quantum}, in quantum thermodynamics \cite{Gemmer,binder2018thermodynamics, myers2022quantum, medina2025}, and also experimentally verified in various settings \cite{batalhao2014experimental, an2015experimental, cerisola2017using, smith2018verification, zhang2018experimental, PhysRevA.100.042119, hernandez2020experimental, Saira2020, PhysRevLett.127.180603, vieira2023exploring}.

Among the experimental platforms, optical tweezers were the first and are still one of the most versatile and widely used. Optical tweezers (OTs) are powerful tools to probe and verify physical phenomena on the microscale \cite{jones2015optical}. These devices use focused laser beams to trap microscopic systems in a noninvasive manner. Using them, researchers can control and analyze systems with high spatial and temporal resolution by tailoring optical potentials \cite{jones2015optical,Pesce2020,gieseler2021optical}. The setup is relatively simple, requiring only standard optical components. 
The precision and simplicity of OTs make optically trapped particles ideal testbeds for experimentally assessing stochastic thermodynamics. Optical trapping has been used in numerous studies to explore these scales \cite{wen2007force,manosas2007force,alemany2012experimental,martinez2013effective,quinto2014microscopic,martins2021dynamically,krishnamurthy2023overcoming,nalupurackal2023towards,Das2023}. For a comprehensive review, including calibration procedures, see Ref.~\cite{gieseler2021optical}.

Despite the importance and significant progress in stochastic thermodynamics over the past decades, key concepts and main results remain underrepresented in standard physics curricula, making this knowledge not widely diffused and sometimes overlooked in their relevance for frontier research in energy, biology, and quantum technologies. 
For newcomers to the field, including researchers and graduate students, another challenge lies in the literature itself, which can be difficult to navigate at first glance. While several excellent reviews have laid out the theoretical foundations \cite{Jarzynski2006, Harris2007, Jarzynski2011, Seifert2012, Seifert2019, Esposito2009, Campisi2011, Benenti2017, Sivak2025}, and landmark experiments have confirmed key predictions across diverse platforms \cite{liphardt2002equilibrium, Wang2002, Carberry2004, collin2005verification, Blickle2006, vanZon2004, garnier2005nonequilibrium, Andrieux2007, douarche2005estimate, koski2013distribution, ciliberto2017experiments, batalhao2014experimental, an2015experimental, cerisola2017using, smith2018verification, zhang2018experimental, PhysRevA.100.042119, hernandez2020experimental, Saira2020, PhysRevLett.127.180603, vieira2023exploring}, a concise hands-on guide integrating theory and experiments for non-specialists is still missing. Most existing works separate abstract theory from the complexities of implementation. Very few detail the experimental procedures, and even fewer share open-source code for calibration and data analysis of raw data.

Here, we aim to close that gap and help update thermodynamics curricula with contemporary verifiable experiments. This paper introduces the main results of stochastic thermodynamics in an accessible and integrated way for students and new researchers. We provide a self-contained, hands-on tutorial that connects theoretical concepts to practical steps for building, calibrating, and analyzing modern experiments using optical tweezers. Specifically, we detail the construction of an OT setup to probe non-equilibrium energy landscapes in real time. By merging theoretical, experimental, and data analysis techniques, we offer practical tools for exploring energy conversion limits and developing efficient micro- and nanoscale devices, potentially aiding the establishment of experimental courses and new research laboratories.

\bigskip
This paper is structured in two main parts: Section \ref{theory} presents the foundations, starting with a brief overview of macroscopic thermodynamics and introducing stochastic thermodynamics and fluctuation theorems. Then it describes the use of optical tweezers to study the thermodynamics of mesoscopic systems. Section \ref{ExperimentalVerification} details the experimental setup, data acquisition and calibration methods, results, and analysis. Finally, Section \ref{Conclusions} provides the discussion and conclusions.

\section{Theory}\label{theory}

This section outlines the necessary theoretical foundations. It begins with a quick review of the first and second laws, establishing the basic concepts and terminology. Next, it introduces the fundamental non-equilibrium work relations, expressed by the Jarzynski equality and the Crooks fluctuation theorem, providing the basic framework and notation needed to understand the analysis in the following sections.

\subsection{Fundamentals of Equilibrium Thermodynamics}

The \textit{first law of thermodynamics} expresses energy conservation for a closed system, distinguishing two mutually exclusive modes of energy transfer: work ($W$) and heat ($Q$). For a system that undergoes an arbitrary process $\Gamma$, the infinitesimal change in internal energy $dU$ is given by \cite{Callen}:
\begin{equation}
 dU = \delta W + \delta Q,
 \label{1law}
\end{equation}
where $\delta W$ and $\delta Q$ represent inexact differentials that depend on the path $\Gamma$ followed by the system. These quantities are fundamentally distinct. Work ($W$) is the energy transferred when external variables (e.g. volume) change under the action of an external agent, while heat ($Q$) is the energy transferred solely due to a temperature difference between systems, via mechanisms such as conduction, convection, or radiation.

The \textit{second law of thermodynamics} introduces the concept of entropy ($S$) and addresses the feasibility and reversibility of processes. Essentially, it constrains possible processes even further, limiting even those that satisfy the first law. For a system in contact with a thermal bath at temperature $T$, it can be expressed as \cite{Callen}:
\begin{equation}\label{2law}
 dS \geq \frac{\delta Q}{T},
\end{equation}
with equality holding for reversible cases. For isolated systems where no heat exchange occurs, the second law simplifies to $dS \geq 0$. For irreversible cases, the entropy change is strictly positive, leading to entropy production.

Equations~\eqref{1law} and~\eqref{2law} are entirely general and can be adapted to more convenient variables by introducing thermodynamic potentials via Legendre transformations. For example, the Helmholtz free energy, defined as $F \coloneqq U - TS$, allows the second law to be reformulated for isothermal (constant temperature) processes as:
\begin{equation}\label{2law2}
 \delta W \geq  dF.
\end{equation}

This form indicates that the minimum (or maximum) work performed (or received) is bounded by the change in free energy between the initial and final equilibrium states, with equality holding only for reversible processes.

\subsection{Brief Overview of Stochastic Thermodynamics} \label{StochasticThermo}

Equilibrium thermodynamics normally ignores fluctuations in macroscopic systems for good reasons. For example, consider a 3D box filled with an ideal monoatomic gas of \(N\) particles in thermal equilibrium at a temperature \(T\). The equipartition theorem gives the average internal energy $\left\langle U \right\rangle_T = \frac{3}{2} N k_B T$, where $\left\langle \boldsymbol{\cdot} \right\rangle_{T}$ represents averages over equilibrium quantities, and \(k_B\) is the Boltzmann constant. The variance of the energy is given by $\text{Var}(U) = \frac{3}{2} N k_B^2 T^2$, and the relative fluctuation in the internal energy,
$\frac{\sigma_U}{\left\langle U \right\rangle_T} = \frac{\sqrt{\text{Var}(U)}}{\left\langle U \right\rangle_T} \approx \frac{1}{\sqrt{N}}$. This relative fluctuation is negligible for a macroscopic system with $N \approx 10^{23}$, and the energy distribution is sharply peaked around its mean value. Allowing fluctuations to be safely ignored in such systems. However, this is no longer true for small systems, such as $N \approx 10^{2}$, when fluctuations become significant compared to the relevant energy scales, requiring them to be included in the thermodynamic description.

Let us consider another illustrative example, a thought experiment in which a box undergoes repeated cycles of the same protocol, such as an isothermal expansion that doubles its volume. In a macroscopic box, where the number of particles is large, fluctuations in pressure and volume due to particle collisions with the walls are negligible. Consequently, the uncertainty in the measured average work is much smaller than the mean work value.

\begin{figure}[tbh]
\centering
\includegraphics[width=6.5cm]{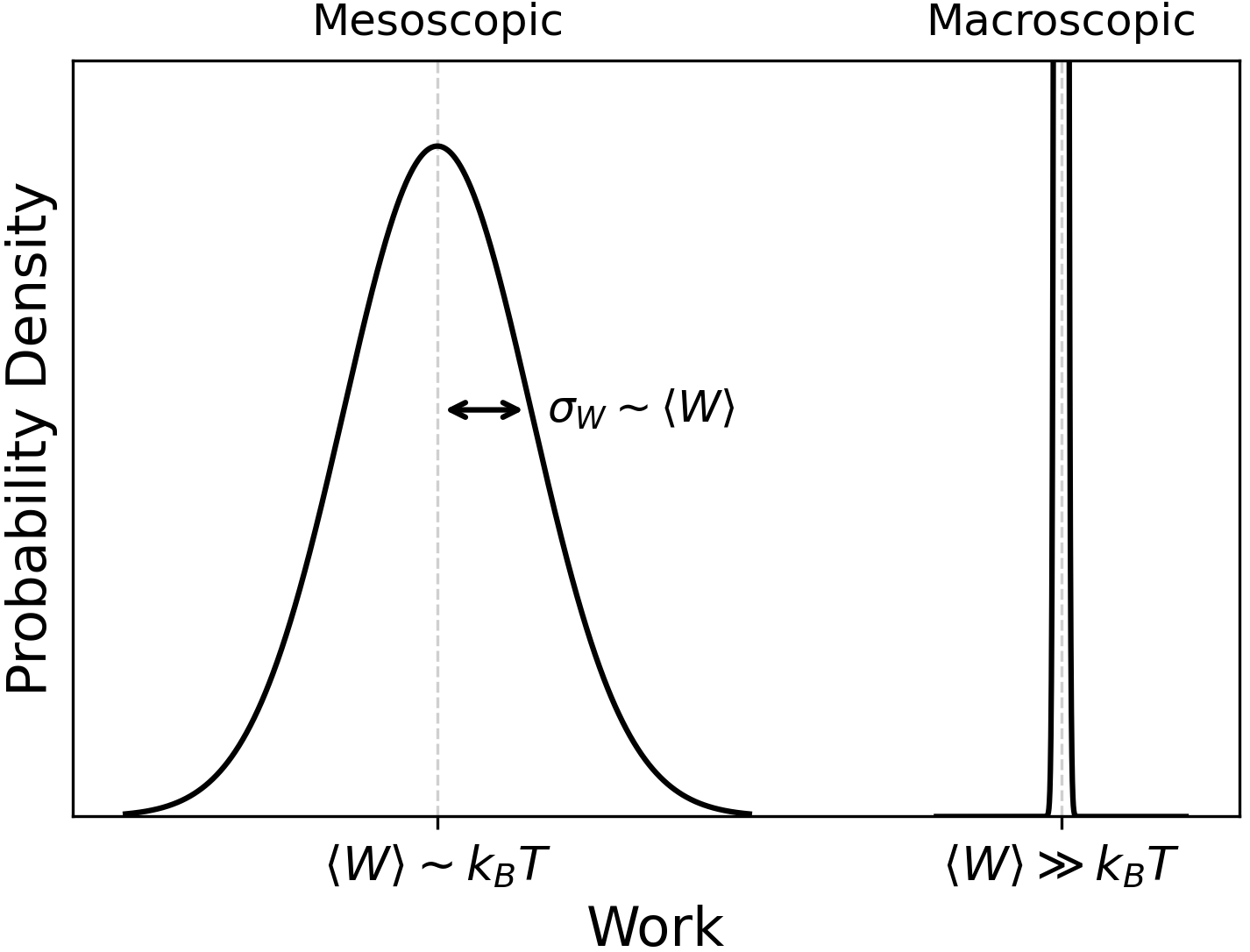}
\caption{Schematic representation comparing the work probability density for mesoscopic and macroscopic systems, with mean values indicated by dashed lines. Fluctuations are negligible in macroscopic systems, and uncertainty is limited only by measurements, while for mesoscopic systems, fluctuations are significant, with values comparable to the average work $\left\langle W\right\rangle$.}
\vspace{-3mm}
\label{fluctuations}
\end{figure}

In contrast, if the same experiment is performed in a microscopically sized box with a small number of particles, the relative fluctuations in pressure become significant, even in an idealized hard-wall box. This leads to noticeable fluctuations in the measured average work, even after many repetitions of the protocol and perfect measurements. Fig.~\ref{fluctuations} illustrates this scenario for mesoscopic systems which are large enough to be readily observable but sufficiently small for fluctuations (on the order of $k_{B}T$) to cause significant measurement deviations.

Stochastic thermodynamics treats quantities like internal energy, work, and heat as stochastic variables, allowing the portrayal of average and individual trajectories in phase space for finite-time processes and non-equilibrium states.

Let $\boldsymbol{x} \coloneqq (\boldsymbol{q}, \boldsymbol{p})$ represent the phase-space coordinates for a small system in contact with a thermal bath, and let $\lambda$ denote an external control parameter (e.g., potential, position, or magnetic field). A modification of $\lambda$, parameterized by time $t$, defines a \textit{protocol}, $\lambda(t)$. Due to the bath's stochastic fluctuations, phase-space trajectories vary even under identical protocols.
The Hamiltonian of the system, $H(\boldsymbol{x}, \lambda)$, represents its internal energy $U(\boldsymbol{x}, \lambda)$. Energy changes can arise from shifts in phase space variables (heat, $Q$) or from changes in external parameters (work, $W$), consistent with the first law of thermodynamics:
\begin{equation}\label{workANDheat}
   dW \coloneqq \frac{\partial U}{\partial \lambda} d\lambda, \qquad dQ \coloneqq \frac{\partial U}{\partial \boldsymbol{x}} d\boldsymbol{x}.
\end{equation}

Therefore, given a protocol $\lambda(t)$, the total work along a single stochastic trajectory over time $t$ is simply
\begin{equation}\label{stochasticWork}
   W=\int_{0}^{t}\frac{\partial U}{\partial\lambda}\frac{d\lambda}{dt^{\prime}}dt^{\prime},
\end{equation}
with an analogous expression for the heat. 

Although it is outside the scope of this introduction, it is worth mentioning that the same reasoning can be applied to define fluctuating entropy \cite{seifert2008stochastic}, for example.

Finally, the ensemble average of work, obtained by repeating the protocol multiple times, is:
\begin{equation}\label{averageWork}
   \left\langle W\right\rangle =\int_{0}^{t}\biggl \langle \frac{\partial U}{\partial\lambda} \biggr \rangle \frac{d\lambda}{dt^{\prime}}dt^{\prime}= \int WP(W)dW,
\end{equation}
where $P(W)$ is the probability distribution function of work and $\langle \cdot \rangle$ represents the ensemble average over all repetions.

\subsubsection{Brownian motion}

The erratic trajectory of a mesoscopic particle immersed in a fluid is well described by the paradigmatic \textit{Brownian motion} \cite{einstein1905motion}. This motion results from collisions with numerous surrounding particles, making it an excellent testbed for the formalism of stochastic thermodynamics. 

Consider a Brownian particle subjected to a one-dimensional potential $U(x)$, initially in thermodynamic equilibrium with its environment (fluid) at temperature $T$. The particle's time evolution is fully characterized by the \textit{Langevin equation} \cite{deGrooth1999,Reichl-book,sekimoto1998langevin}:
\begin{equation}\label{Langevin}
  m \ddot{x} = -\gamma \dot{x} -\frac{d U(x)}{d x} + F_{th},
\end{equation}
where $F_{th}$ is the stochastic force, given by Gaussian white noise arising from the system's interaction with its thermal reservoir characterized by: $\langle F_{th} (t)\rangle = 0$ and $\langle F_{th} (t)F_{th} (t')\rangle = 2\gamma k_B T \delta (t-t')$, with $\delta$ being the Dirac delta function. 
In Eq.~(\ref{Langevin}), $m$ is the mass of the particle, $\gamma=6\pi \eta R$ denotes the particle's friction coefficient, where \( \eta \) is the medium's viscosity and \( R \) the particle's radius. Thus, the particle's dynamics are governed by the deterministic force from the external potential and the stochastic thermal forces, which are non-negligible at the mesoscopic scale. 

\begin{figure}[tbh]
\centering
\includegraphics[width=6.5cm]{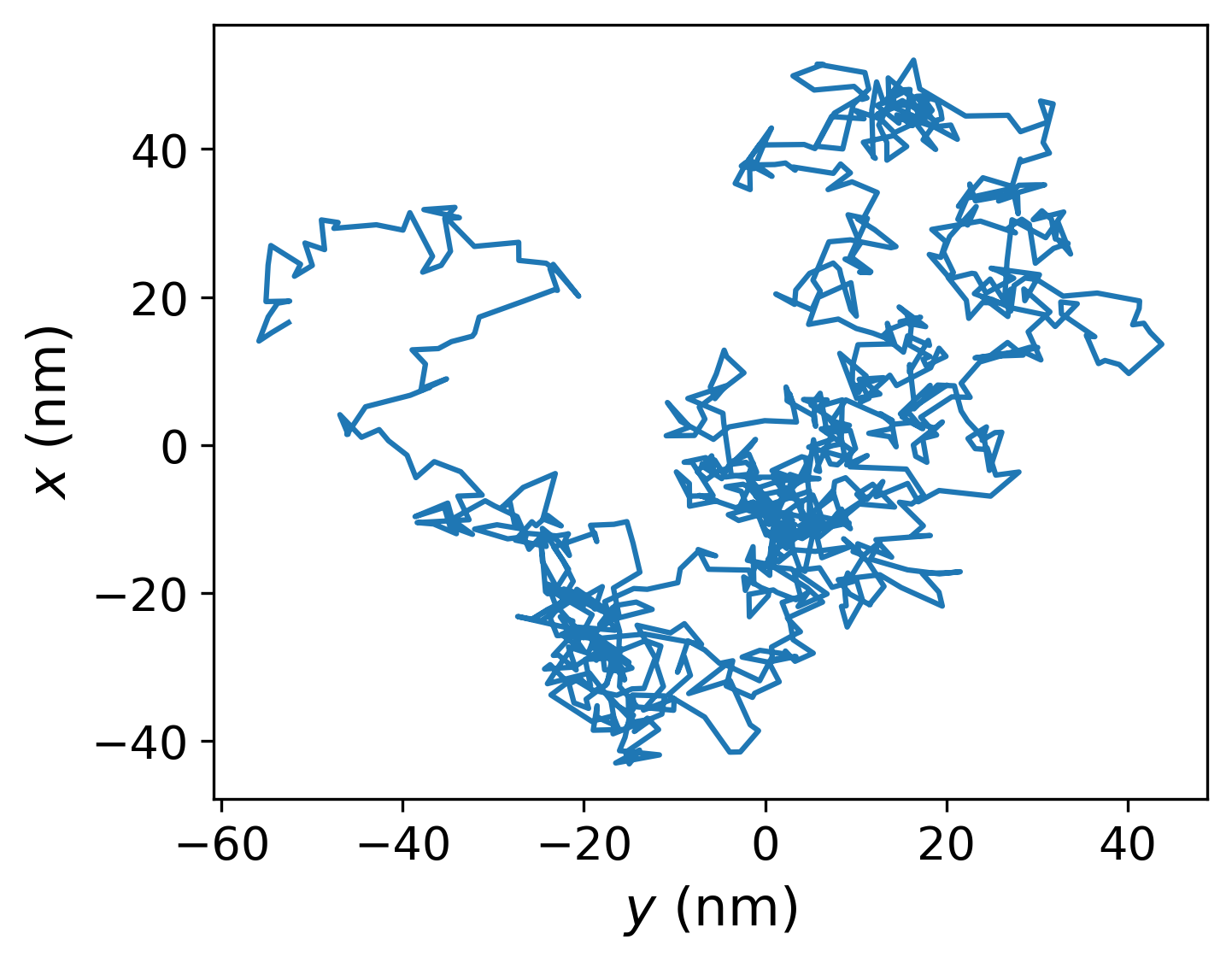}
\includegraphics[width=6.5cm]{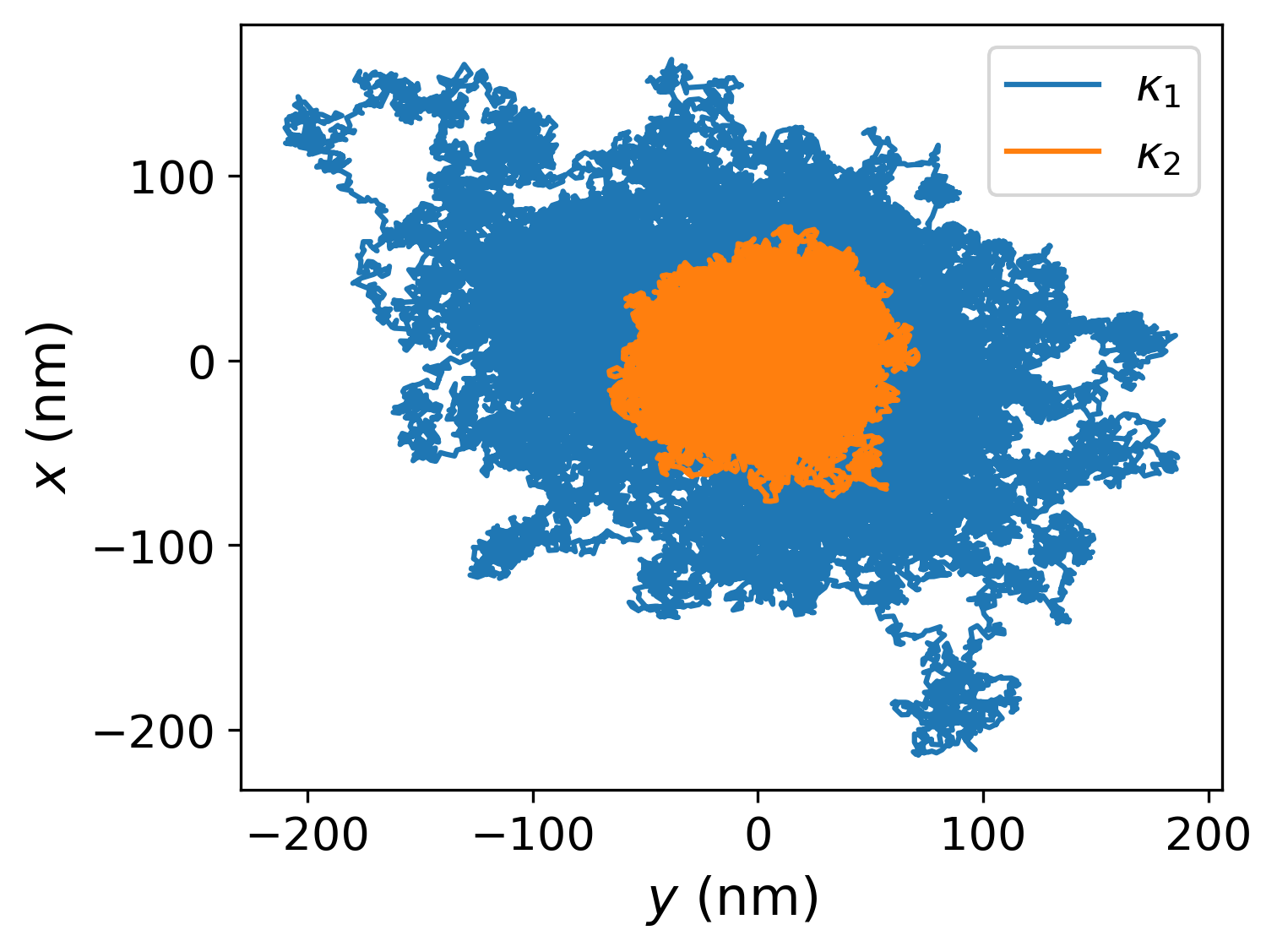}
\vspace{-3mm}
\caption{Simulated trajectories of a water-immersed particle trapped at room temperature in a harmonic potential. (Top) Trajectories for a trap stiffness \( \kappa_1 = 1 \ \mathrm{pN/\mu m} \) over \( 10 \ \mathrm{ms} \). (Bottom) Comparison of trajectories for trap stiffnesses \( \kappa_1 \) and \( \kappa_2 = 10 \ \kappa_1 \) over \( 1 \ \mathrm{s} \). The simulations use a time interval of \( 10 \ \mathrm{\mu s} \) between points with parameters: \( T = 300 \ \mathrm{K} \), \( \gamma = 6\pi \eta R \), \( \eta = 0.001 \ \mathrm{N \cdot s \cdot m^{-2}} \), and \( R = 1 \ \mathrm{\mu m} \).}
\label{brownianmotionharmonic}
\end{figure}

Figure~\ref{brownianmotionharmonic} shows a simulation \cite{volpe2013simulation} of the Brownian motion for a particle immersed in water in a harmonic potential \( U(x) = \frac{1}{2} \kappa x^{2} \) over a time interval of \( \tau = 10 \ \mathrm{ms} \), in the upper part. The lower part of the figure contrasts this behavior with \( \tau = 1 \ \mathrm{s} \) for different trap stiffnesses, \( \kappa_1 = 1 \ \mathrm{pN/\mu m} \), and \( \kappa_2 = 10\, \kappa_1 \). Higher values of the force constant \( \kappa \) result in the particle being more spatially constrained, as is evident from the smaller standard deviation of its center of mass position.

Sekimoto \cite{sekimoto1998langevin,Sekimoto-book} showed that when the particle is subjected to an external control parameter $\lambda(t)$, the trajectory can be directly related to thermodynamic quantities such as heat and work. To establish this connection, consider the overdamped regime of the Langevin equation, where the inertial term $m \ddot{x}$ is negligible. After rearranging the terms, we obtain:
\begin{equation}
    \gamma \dot{x} = -\frac{d U(x)}{d x} + F_{th}(t).
\end{equation}

According to Sekimoto \cite{sekimoto1998langevin}, the heat  $d \mathcal{Q}$ exchanged with the thermal reservoir is defined as:
\begin{equation}
    d \mathcal{Q} = -\left( -\gamma \dot{x} + F_{th}(t) \right) dx,
\end{equation}
where the sign convention here is opposite to the usual in macroscopic thermodynamics, therefore $d \mathcal{Q} = - dQ$.

Using the first law of thermodynamics, $dU = dQ + dW$, where $dW$ is the work done on the system, we can write:
\begin{equation}
    dW = d \mathcal{Q} + dU = \frac{\partial U}{\partial \lambda} d\lambda.
\end{equation}

Thus, the stochastic work applied to the system by the external control along the trajectory $x(t)$ is given by Eq.~\eqref{stochasticWork}, and the stochastic heat dissipated by the system into the bath is:
\begin{equation}
    \mathcal{Q} = \int_{0}^{t} dt^{\prime} \left[ -(-\gamma \dot{x} + F_{th}(t^{\prime})) \dot{x} \right].
    \label{eq:heatcili}
\end{equation}
Therefore, from the particle's trajectory $x(t)$ and values of the control parameter $\lambda(t)$ of the modulated potential, it is possible to compute the time evolution of work and heat during a given (single) process and determine their final values using Eq.~(\ref{stochasticWork}) and Eq.~(\ref{eq:heatcili}). Due to the stochastic force, each realization will result in different values, and one can obtain their average across an ensemble of trajectories, as in Eq.~(\ref{averageWork}).

For completeness, one might be interested in an ensemble-level description of this physical setting, rather than the individual trajectory picture provided by the Langevin equation. This is captured by the \textit{Fokker-Planck equation} \cite{siegman1979simplified,Kamizaki2022}, which gives the dynamics of the probability density function $P(x,t)$ corresponding to the particle's position $x$ at time $t$:
\begin{equation}
  \frac{\partial P(x,t)}{\partial t} = \frac{1}{\gamma}\frac{\partial}{\partial x} \left[\frac{d U(x)}{d x} P(x,t) + k_B T \frac{\partial P(x,t)}{\partial x}\right]. 
  \label{eq:FokkerPlanck}
\end{equation}

\subsection{Fluctuation theorems}\label{FlucTheorems}

To accurately describe the thermodynamics of mesoscopic systems, one cannot neglect the role of fluctuations. Surprisingly, these fluctuations carry valuable information. The FTs provide formal mathematical relations that express certain system symmetries, usually expressed in terms of the probability distribution of some stochastic variable. This is a broad and rich subject, but here we will focus on two emblematic results that helped spur the entire field of study. The reader can find a more comprehensive discussion in \cite{Harris2007,seifert2008stochastic,ritort2008nonequilibrium, Esposito2009,schuster2013nonequilibrium,Peliti-book,Seifert-book}. In the following, we introduce the non-equilibrium work relations \cite{Jarzynski1997prl,crooks1998nonequilibrium,crooks1999entropy,Jarzynski2011}.

\subsubsection{Jarzynski equality}

One of the most relevant examples of FT is the Jarzynski equality (JE), proposed in \cite{Jarzynski1997prl} by Christopher Jarzynski. It relates the non-equilibrium work distribution within a given protocol to the difference in Helmholtz free energy, $\Delta F$. 

Let us assume our thermodynamic system of interest is described by a Hamiltonian $H(\boldsymbol{x},\lambda)$, where $\lambda$ represents the external control parameter. Now, consider the following simple procedure: \textbf{(i)} Initially, place the system in thermal contact with a reservoir at temperature $T$, with $\lambda$ fixed at $\lambda_i$; \textbf{(ii)} After equilibration, modify $\lambda$ according to some predefined finite-time protocol until it reaches a final value\footnote{For simplicity, the system and reservoir were assumed to be decoupled during protocol execution. However, in the proposed experiment, the bath remains coupled to the system throughout the process. Although there are subtleties when dealing with \textit{strong-coupling} scenarios (see \cite{Jarzynski2011} and \cite{ChrisJarzynski_2004} for more details), Eq.~\eqref{jarzynski} remains valid for both cases.} $\lambda_f$. During this execution, stochastic work is performed, as quantified by Eq.~\eqref{stochasticWork}. After repeating this process several times, one can calculate the statistics for this process. Under these conditions, the following equality holds:
\begin{equation}\label{jarzynski}
    \left\langle e^{- W/k_{B}T}\right\rangle =e^{-\Delta F/k_{B}T},
\end{equation}
where $\Delta F \coloneqq  F_f - F_i$, with $F_{i,f}$ being the Helmholtz free energy relative to the equilibrium state characterized by $\lambda_{i,f}$. 

This expression is surprising for several reasons. First, it is an equality, which contrasts with the usual thermodynamic relation between work and free energy for non-equilibrium processes (Eq.~\eqref{2law2}). Additionally, its derivation and validity do \textit{not} depend on the specific protocol or the system's final state, even though the latter strongly depends on the former. This means that as long as the system reaches $\lambda_f$, the work protocol can be executed as rapidly as desired, potentially driving the system to states far from equilibrium. Its main relevance is that one can gain access to helpful equilibrium quantities without relying on quasistatic processes by simply collecting work statistics from arbitrary finite-time processes.

Furthermore, applying Jensen's inequality $\left\langle e^{x}\right\rangle \geq e^{\left\langle x\right\rangle}$ to Eq.~\eqref{jarzynski}, it is possible to recover the standard statement of the second law (Eq.~\eqref{2law2}): 
\begin{equation}
 \left\langle W \right\rangle \geq  \Delta F.
\end{equation}

Notice that this inequality refers to the average value of work, indicating that individual stochastic trajectories may eventually exhibit work values below $\Delta F$.

\subsubsection{Crooks fluctuation theorem}

Following Jarzynski's proposal \cite{Jarzynski1997prl}, Gavin Crooks introduced another general equality related to similar quantities in \cite{crooks1998nonequilibrium,crooks1999entropy}. 

To understand Crooks fluctuation theorem (CFT), consider the same basic procedure as before, but now, both ways are needed: forward and backward. Define a \textit{forward-protocol} as a sequence where $\lambda$ changes from $\lambda_i$ to $\lambda_f$, and a \textit{backward-protocol} as the time-reversed sequence where $\lambda$ changes from $\lambda_f$ back to $\lambda_i$. In forward (backward) execution, the system starts in equilibrium according to $\lambda_{i(f)}$. 

The CFT is expressed as:
\begin{equation}
    \frac{P_F(W)}{P_R(-W)}=e^{(W-\Delta F)/k_B T},
    \label{eq:Crooks}
\end{equation}
where $P_F(W)$ is the probability of performing the work $W$ during the execution of the forward protocol, while $P_R(-W)$ is the probability of attaining the work $-W$ during the execution of the backward (reverse) protocol. 

This expression encodes several essential insights. First, notice that the distributions $P_F(W)$ and $P_R(-W)$ coincide exactly when $W=\Delta F$. This feature allows the straightforward identification of $\Delta F$ by simply observing the intersection point of both stochastic work distributions. Additionally, by directly integrating Eq.~\eqref{eq:Crooks} and using the normalization of the probability distributions $\int_{-\infty}^\infty dW\, P_R(-W) = 1$, one can easily derive the Jarzynski equality in Eq.~\eqref{jarzynski}.

Thus, the Crooks fluctuation theorem provides a deeper understanding of non-equilibrium processes and connects forward and reverse protocols, reinforcing the fundamental principles of thermodynamics at the mesoscopic scale.

\subsection{Probing stochastic thermodynamics with optical tweezers}

\subsubsection{Optical tweezers}

The development of the first laser optical traps began in the late 1960s, when Arthur Ashkin observed that a laser beam could push micrometer-sized transparent particles due to radiation pressure. Subsequently, the counter-propagating beam trap was used to demonstrate the first all-optical trap \cite{ashkin1970acceleration}, and years later, the first single-beam trap, known as \textit{optical tweezers}, was introduced \cite{ashkin1986observation}. This tool revolutionized the way scientists interact with microscopic objects, mainly due to its ability to precisely manipulate micro and nanoscale systems in a noninvasive manner using forces exerted by light. In fact, due to its numerous applications in various scientific fields, ranging from studying biological systems \cite{ashkin1990force, block1990bead, bustamante1994entropic, finer1994single} and chemical processes \cite{yi2006microfluidics} to studies of thermodynamics at the microscale \cite{liphardt2002equilibrium, Wang2002, collin2005verification, Blickle2006, toyabe2010experimental, berut2012experimental}, optical trapping has become an essential modern tool. Arthur Ashkin was awarded the 2018 Nobel Prize in Physics for developing this powerful technique.

\begin{figure}[tbh]
\centering
\includegraphics[width=7.6cm]{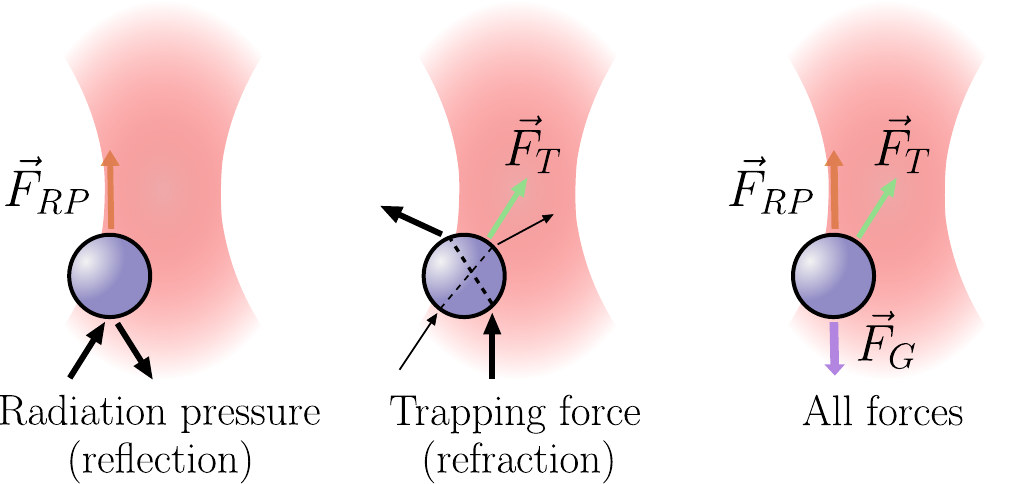}
\caption{The basic scheme of the beam's trajectory (black lines) during the reflection and refraction, and the resulting forces are radiation pressure (orange) and trapping force (green), respectively. The gravitational force (purple) is also represented but usually is much smaller.}
\label{fig:OTrays}
\end{figure}

For the discussion here, focused on stochastic thermodynamics, and given the size of the optically trapped particle (with $R \approx 1 \, \mu$m), we can safely use the simplest description for the operation of the optical tweezer, using geometric optics and momentum conservation. For further details and more general descriptions, we refer to Refs.~\cite{jones2015optical,Pesce2020, gieseler2021optical, martins2024thesis}.

An optical tweezer employs a highly focused laser beam to manipulate a microscopic particle (Fig.~\ref{fig:OTrays}). As light reflects and refracts upon entering and exiting the particle, its momentum changes, generating a radiation pressure force ($\vec{F}_{RP}$) along the beam axis and a trapping (gradient) force ($\vec{F}_T$) toward the region of highest intensity at the focal point. For materials with an index of refraction larger than its surroundings, like silica beads in water, $\vec{F}_T$ is attractive, balancing $\vec{F}_{RP}$ and the small gravitational force ($\vec{F}_G$) to achieve stable trapping. 

The resulting force can be approximated as a linear restoring force for small displacements of particles from their equilibrium position, expressed as \( F_{OT} = -\kappa x \). In this regime, the trapped particle behaves as a mass-spring system in a harmonic potential $U = \frac{1}{2} \kappa x^2$, as illustrated in Fig.~\ref{potential}. In addition, it is possible to dynamically control relevant parameters such as the spring constant $\kappa$ and the equilibrium position simply by adjusting the intensity and position of the beam. Moreover, by appropriately controlling the temporal and spatial modulations of the beam, a wide variety of optical potentials can be generated \cite{albay2018optical, martins2021dynamically, Kamizaki2022b}, making OTs a versatile platform for experimental research.

\subsubsection{Mechanical analogy \& control protocols}
\label{BreathingParabola}

In classical thermodynamics, the compression and expansion of a piston containing an ideal gas provide one of the simplest scenarios for the theory. Starting from equilibrium, external work is required to compress the piston into a smaller volume, while work is produced during the reverse process. In this case, the control parameter $\lambda(t)$ is simply the position of the piston as a function of time.

\begin{figure}[tbh]
\centering
\includegraphics[width=7cm]{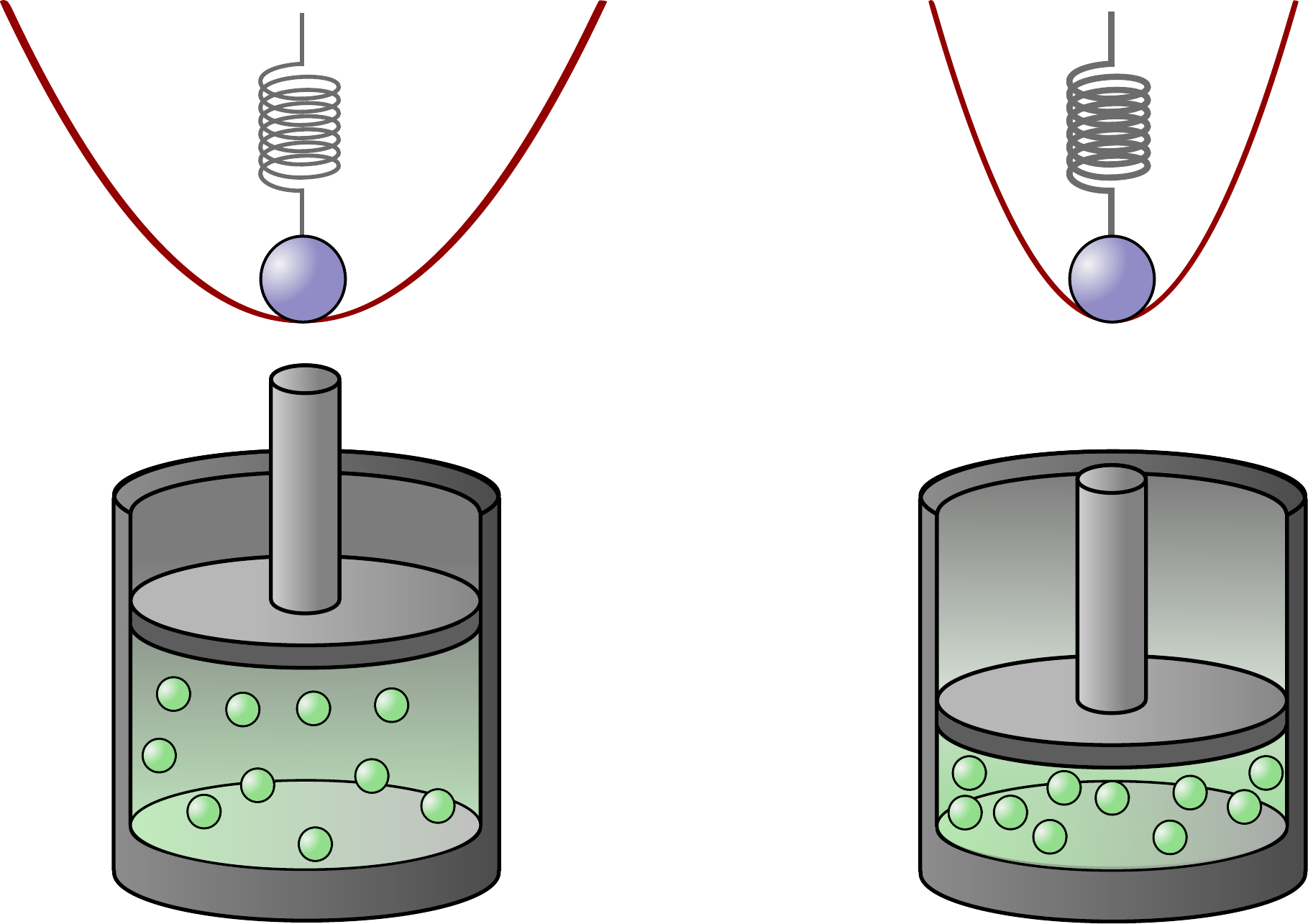}
\caption{A schematic representation of a piston showing two different volumes, analogous to a particle trapped in a harmonic potential with two corresponding trap stiffnesses.}
\label{fig:piston}
\end{figure}

This emblematic macroscopic scenario maps directly onto the mesoscopic regime using an optically trapped microsphere as a prototype model for a Brownian particle in a harmonic potential, as illustrated in Fig. \ref{fig:piston}. In this analogy, the role of the piston is played by a dynamically controlled optical potential $U(x,t)=\frac{1}{2}\kappa(t) x^2$, where the control parameter is the trap stiffness of the optical potential. Thus, from now on, the external protocol parameter corresponds to $\lambda(t)\equiv\kappa(t)$.

\begin{figure}[tbh]
\centering
\includegraphics[width=7cm]{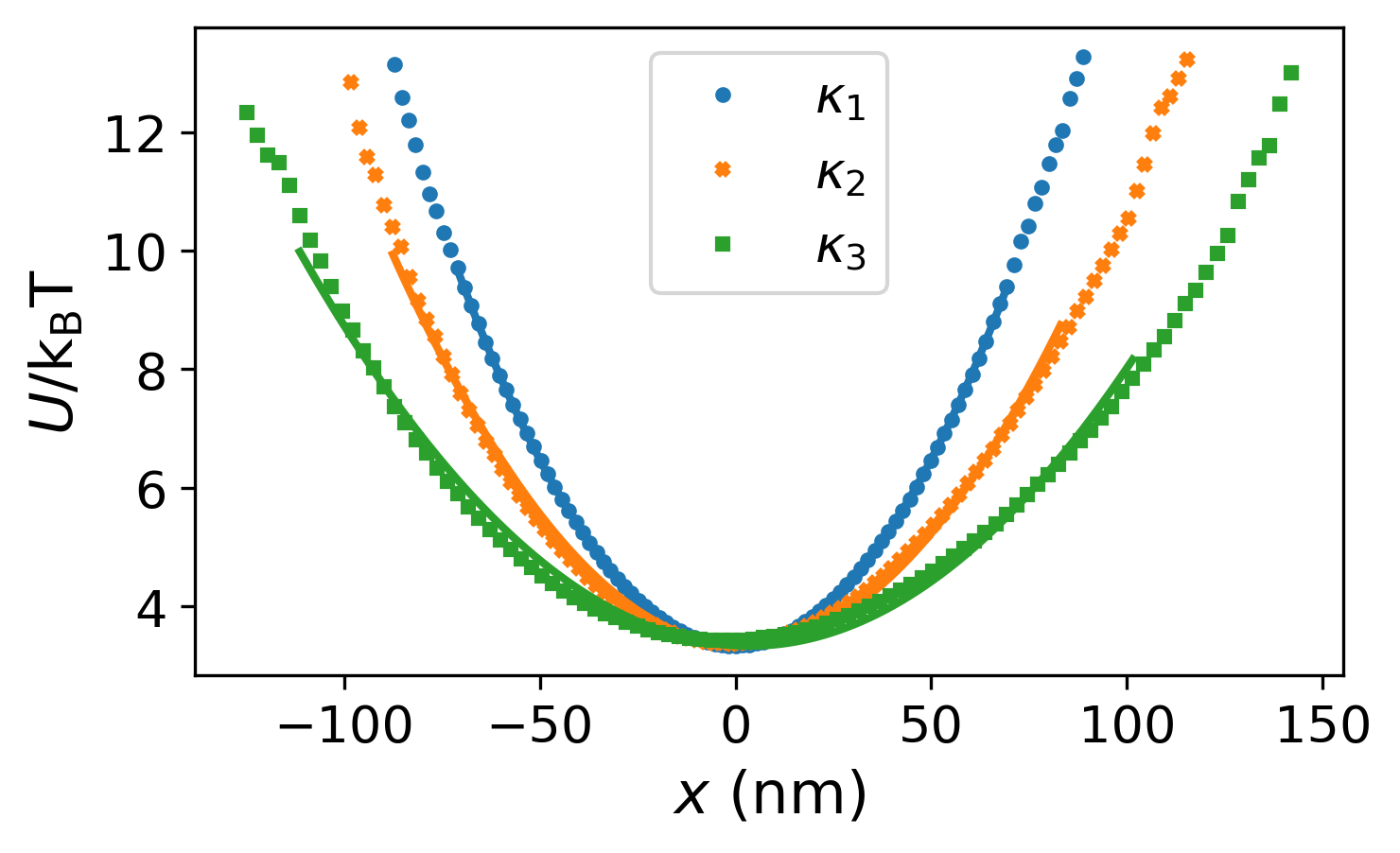}
\vspace{-2mm}
\caption{Measured harmonic potentials for different force constants $\kappa$ (acting as the control parameter $\lambda$), with $\kappa_{1}>\kappa_{2}>\kappa_{3}$. Results were obtained from position histograms of a trapped $2 \ \mathrm{\mu m}$ silica bead. The value of $\kappa$ is proportional to the laser intensity. In experiments, the compression and expansion protocols are executed by modulating the intensity.}
\label{potential}
\end{figure}

In the \textit{compression process}, the stiffness of the trap increases, restricting the particle to a smaller region. This is analogous to the piston case. In contrast, as the stiffness decreases, the trap becomes weaker, allowing the particle to access a larger volume, representing the \textit{expansion process}. Figure \ref{potential} shows this connection explicitly with potentials reconstructed directly from experiments at different trap strengths. To calculate the potential directly from the particle position data, one starts by allowing thermal equilibrium with the environment (fluid) at temperature $T$. Thus, its probability distribution follows the Maxwell-Boltzmann distribution.
\begin{equation}
    P(x) = P_0 \exp\Big[-\frac{U(x)}{k_B T}\Big],
    \label{Eq1}
\end{equation}
where $P_0$ is a normalization factor, such that $\int P(x) dx = 1 $. 
Hence, the potential $U(x)$ can be readily determined by
\begin{equation}
    U(x) = - k_B T \ln[P(x)]+U_0,
    \label{eq:potencial1}
\end{equation}
where $ U_0 $ is an arbitrary additive constant. 

This simple experiment provides a quick and essentially model-independent outline of the optical potential, which is very practical in the lab but usually does not have enough precision. In section~\ref{calibration}, we present a more precise and robust calibration method for the case of a harmonic potential.

Since the particle remains immersed in a fluid with fixed temperature $T$, performing isothermal processes is straightforward. The control protocol $\lambda(t)$ (corresponding to stiffness) can be easily programmed for different types of profiles, allowing for various experiments. For simplicity, here, we focus on a linear protocol:
\begin{equation}
    \lambda (t) = \lambda_i +\frac{\Delta \lambda}{\tau_P} t,
    \label{eq:linearprotU}
\end{equation}
where $\Delta \lambda = \lambda_f-\lambda_i$ is the modulation amplitude and $\tau_P$ is the time duration of the protocol.

The particle's Hamiltonian is given by $H(t) = p^{2}/2m+U(t)$, where $p$ is the particle's momentum, and $U(t)$ is the trapping harmonic potential
\begin{equation}\label{harmonicPotential}
    U(t) = \frac{\lambda(t)x^2}{2}.
\end{equation}

Thus, the partition function for a given $\lambda$ is
\begin{equation}
Z=\int_{-\infty}^{+\infty}dx\,\int_{-\infty}^{+\infty} dp\,e^{-H/k_{B}T}=Z_{p}\sqrt{\frac{2\pi k_{B}T}{\lambda}},
\end{equation}
where $Z_{p}=\int_{-\infty}^{+\infty}dp\,e^{-p^{2}/2mk_B T}=\sqrt{2\pi m k_B T}$.

For a finite-time process, the work relative to a single run of the protocol  can be computed using Eqs.~\eqref{stochasticWork}, \eqref{eq:linearprotU}, and \eqref{harmonicPotential}: 
\begin{equation}
     W=\frac{\Delta\lambda}{2\tau_{P}}\int_{0}^{\tau_{P}}x^{2}(t)\,dt.
     \label{eq:workused}
\end{equation}

For an ensemble of trajectories, the average work is
\begin{equation}
    \langle W\rangle=\int_{0}^{\tau_{P}}dt\frac{d\lambda(t)}{dt}\left\langle \frac{\partial U(x,\lambda)}{\partial\lambda}\right\rangle =\frac{\Delta\lambda}{2\tau_{P}}\int_{0}^{\tau_{P}}\langle x^{2}(t)\rangle\,dt.
     \label{eq:workaverage}
\end{equation}
From Eq.~\eqref{eq:FokkerPlanck} and the initial condition obtained from the equipartition theorem of energy, i.e., $\langle x^2(0) \rangle =k_B T/\lambda_i$, it is possible to estimate $\langle{x^2}\rangle$ and compute the average work. 

Finally, given the Helmholtz free energy $F\coloneqq -k_B T \ln Z$, one can obtain $\Delta F$ relative to the equilibrium states characterized by $\lambda_i$ and $\lambda_f$:
\begin{equation}
\Delta F  = - k_B T \ \ln\frac{Z_f}{Z_i} = - k_B T \ \ln \sqrt{\frac{\lambda_i}{\lambda_f}}.
\label{eq:deltafeq}
\end{equation}

Note that, given Eq.~\eqref{2law2} for the second law, one can quickly approximate this quantity as the minimum amount of energy necessary to pay for driving the particle's state from $\lambda_{i}$ to $\lambda_{f}$, such that $\left\langle W\right\rangle \geq \Delta F$, where the equality is only achieved by infinite-time protocols (quasi-statically).

In conclusion, it is possible to gather statistics that recover meaningful thermodynamic quantities by measuring the particle's position during many repetitions of the same protocol. This can be done at the individual trajectory levels, by examining how work evolves during a single protocol run, and at the ensemble level, by repeating the process N times. However, a critical experimental challenge lies in accurately characterizing and controlling the external potential that the particle experiences, while also accurately recording its position.

\section{Experiment}\label{ExperimentalVerification}

This section presents the experimental part of this work, demonstrating how to verify the fluctuation theorems using a custom-built optical tweezers setup. First, we introduce the experimental configuration and outline the calibration process for characterizing the harmonic potential. Then, we present the results and their analysis. 

In short, the physical system under investigation is a trapped $2\ \mathrm{\mu m}$ silica microsphere (bead) manipulated dynamically using optical fields. Controlling the optical potential while simultaneously recording the bead's position, we can calculate the stochastic trajectories introduced in section~\ref{StochasticThermo} and use the gathered statistics of the ensemble to verify Eqs.~\eqref{jarzynski} and~\eqref{eq:Crooks}.

\subsection{Experimental setup}

\begin{figure}[tbh]
\centering
\includegraphics[width=7.5cm]{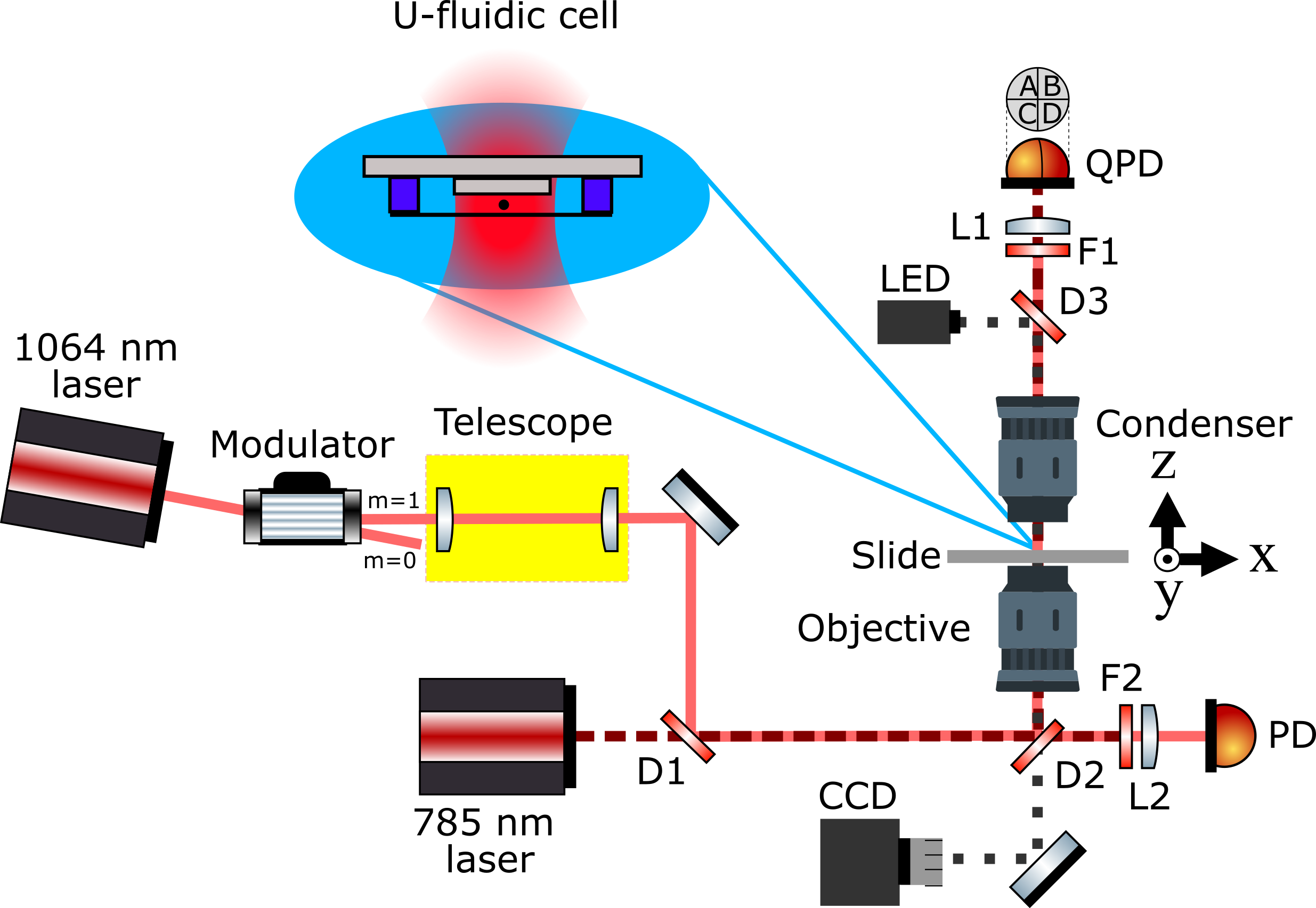}
\caption{The experimental setup consists of a trapping system with a $1064 \ \mathrm{nm}$ infrared laser passing through a 2-axis acousto-optic deflector (AOD). A telescope expands the beam waist to fill the entrance of a $63\times$ immersion objective lens ($\mathrm{NA}=1.32$, Leica Germany HCX PL APO). For position detection, a $785 \ \mathrm{nm}$ laser is aligned with the trapping laser. A condenser collects the scattered light and directs it through lens L1, which reduces the beam's diameter to fit within the active area of the Quadrant Photodiode (QPD). Additionally, a white LED and a CCD camera (Mikrotron MC 1310) are also used to visualize and record the particle's position. Mirrors and dichroic filters (D1, D2, D3) guide the beams. A notch filter (F1) blocks all wavelengths except the $785 \ \mathrm{nm}$ light, allowing it to reach the QPD, while a long-pass filter (F2) prevents the $785 \ \mathrm{nm}$ light from reaching the photodetector (PD) used to monitor the trapping IR laser. Lens L2 focuses the infrared beam onto the PD to maximize its signal. The inset at the top depicts the custom-designed U-shaped fluidic cell that enables long-duration measurements by preventing interference from other particles.}
\label{fig:setupENS}
\end{figure}

The experimental setup is illustrated in Figure \ref{fig:setupENS}. A 1064 nm infrared (IR) laser (FORTE 02163 1064-SLM, supply 93580 LD 3000, Laser Quantum) is used to trap the particles. The laser is connected to two perpendicular acousto-optic deflectors (AODs), specifically the Optoelectronic DTSXY model, which allows precise control over the beam's intensity and angular position.

The 2-axis AOD uses piezoelectric transducers to modulate an optical medium (TeO$_2$ crystal). When a radio frequency (RF) signal from a function generator (AFG3102) drives the piezoelectric device, an acoustic wave propagates through the crystal, diffracting the passing light. By aligning the system with the first order of diffraction (m=1), we gain control over the intensity and angular position of the beam by adjusting the amplitude and frequency of the RF signal. The two perpendicular AODs enable independent control of the beam's position along the $x$ and $y$ axes; however, we will focus only on the intensity modulation for the proposed protocols ($\lambda(t)$) and movements along one direction ($x(t)$).

A $63\times$ immersion objective lens ($\mathrm{NA}=1.32$, Leica Germany HCX PL APO) focuses the IR beam to trap the particle. A telescope composed of two lenses collimates and magnifies the trapping laser, ensuring that it fills the entrance pupil of the objective. This maximizes trapping efficiency, as the gradient force, which is proportional to the intensity gradient, is optimized by tightly focusing the beam. After the objective, a sample holder connected to a 3D nanopositioning stage allows for precise adjustments of the sample's position. We trap low-concentration $2 \ \mathrm{\mu m}$ silica beads within a U-fluidic cell (see inset in Fig.~\ref{fig:setupENS}). This homemade cell is constructed by gluing a slide to a coverslip using double-sided tape and attaching a glass cylinder in the center to minimize the distance between the surfaces to approximately $50 \ \mathrm{\mu m}$. This design can significantly reduce the influx of particles in the measurement region, facilitating long-term data acquisition \cite{berut2015thesis,martins2024thesis}. For measurements, the monitored particle is captured at the edges of this region and optically dragged into the central area.  

For particle visualization, we use a bright-field CCD camera (Mikrotron MC 1310) and a white LED for illumination. The trapped particles are tracked using a Quadrant Photodiode Detector (QPD) to achieve the high sampling rate necessary for out-of-equilibrium processes. A low-power $785 \ \mathrm{nm}$ laser (Thorlabs LP785-SF20) is integrated into the trapping path and aligned collinearly with the trapping laser. The interference between the incoming and scattered fields produces a position-dependent pattern, collected by a condenser with a numerical aperture $\mathrm{NA}=0.53$ (Leica 521500) and directed to a homemade QPD. The 4 QPD output voltage signals ($A$, $B$, $C$, and $D$), from which the particle's center of mass positions are calculated as:
\begin{align}
    x_{\mathrm{QPD}} &= (A + C) - (B + D), \\
    y_{\mathrm{QPD}} &= (A + B) - (C + D).
\end{align}

Finally, the control of the function generator, along with readings from the QPD voltages and the PDA36A-EC photodetector, which monitors the power of the infrared beam, is managed using a high-speed, multifunction data acquisition (DAQ) board (National Instruments PXI-4472).

\subsection{Characterizing harmonic potentials} \label{calibration}

Various methods exist to determine the trap stiffness, $\kappa_{x(y)}$, and the factor $S_{x(y)}$, which relates the QPD output to actual position readings. Section~\ref{BreathingParabola}, in Eq.(\ref{eq:potencial1}), discussed one of the simplest. However, the power spectral density (PSD) analysis \cite{Berg-Sørensen2004} is the most reliable approach \cite{jones2015optical}, and it is the method used in this study.
The procedure involves trapping a particle in a static potential and recording the position $x(t)$ at a sampling frequency $f_s$ over a time interval $T_s$. The analysis focuses on the $x$-axis, but the same applies to the $y$ and $z$-axes. 

From the position data $x(t)$, a numerical discrete Fourier transform can convert it to the frequency domain, yielding $\breve{x}(f)$. The PSD is then calculated as:
\begin{equation}
    P(f) = \lim_{T_s\rightarrow+\infty} \frac{|\breve{x}(f)|^2}{T_s}.
    \label{eq:normalPSD}
\end{equation}

\begin{figure}[tbh]
\centering
\includegraphics[width=8cm]{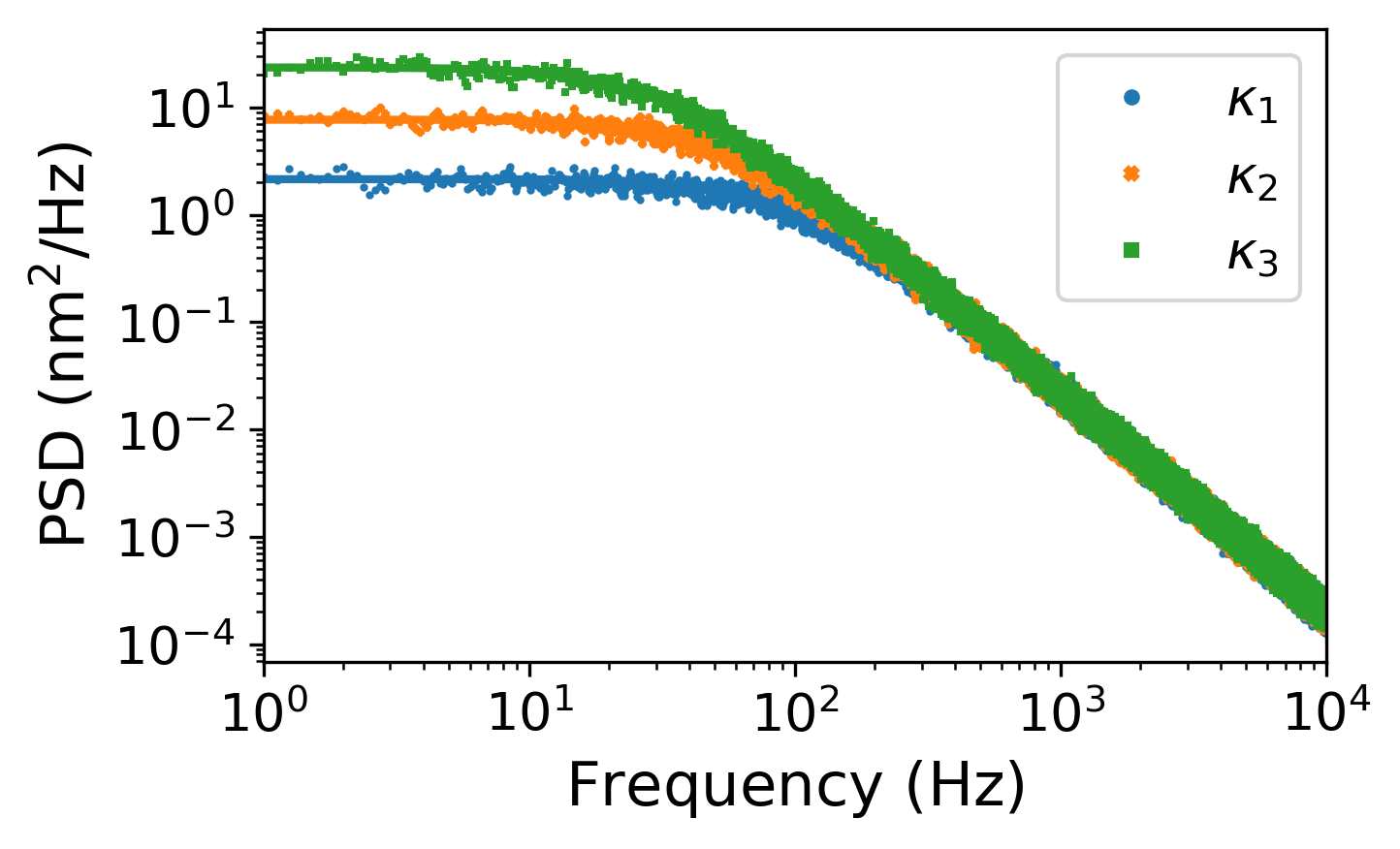}
\caption{Power spectra density (PSD) analysis of a $2\ \mathrm{\mu m}$ silica bead trapped at three different laser powers, increasing from green to blue ($\kappa_{1}>\kappa_{2}>\kappa_{3}$). Higher laser power increases trap stiffness and lowers the PSD baseline plateau at low frequencies.}
\label{fig:PSD}
\end{figure}

For PSD computation, we used Welch's method \cite{welch1967use}, using an acquisition rate at $f_s=100 \ \mathrm{kHz}$ and $T_s=30 \ \mathrm{s}$. This resulted in the number of points $N_{\mathrm{FFT}} = 8 \times 10^{5}$ used for characterization of $\kappa_x$ and $S_x$, with an overlap of $90\%$ between segments. Figure \ref{fig:PSD} displays the PSD from experimental data of a $2\ \mathrm{\mu m}$ silica bead trapped at three different laser powers. Higher laser power corresponds to increased trap stiffness and reduced mean-squared displacement within the trap. At each power level, $10$ measurements were taken and the PSD presented represents the average of these measurements. Laser power is controlled directly via the AOD amplitude channel. 

The analytical PSD for a harmonic trap in the overdamped regime is given by the Lorentzian form \cite{jones2015optical, Berg-Sørensen2004}:
\begin{equation}
    P(f) = \frac{S_x^2 D}{2\pi^2 (f^2 + f_c^2)},
    \label{eq:normalPSD_corrected}
\end{equation}
where $D = k_B T/\gamma$ is the diffusion coefficient, $S_x$ is the amplification factor converting the QPD voltage output to position ($x(t)=x_{\mathrm{QPD}}(t)/S_x$), $f_c = \kappa_x/2 \pi \gamma$ is the corner frequency, and $\gamma = 6\pi \eta R$ is the particle's friction coefficient ($\eta$ is the medium viscosity and $R$ is the radius of the particles). Detailed derivations are given in~\cite{martins2024thesis}. From data and using the least squares fit \footnote{While Eq.~(\ref{eq:normalPSD_corrected}) is the standard method for calibrating optical tweezers, we use a fitting function that also takes into account hydrodynamic interactions and inertial effects, as detailed in \cite{franosch2011resonances}. According to \cite{lukic2007motion}, memory effects become significant for trap calibration at high sampling frequencies, making these corrections essential for accurate estimations of potential. Although not strictly required, these corrections may improve calibration accuracy. In the range used here, we find approximately $1–4\%$ differences in the amplification coefficient and $5–10\%$ in the trap stiffness.}, one can extract the trap stiffness $\kappa_x$ and the amplification factor $S_x$.
 
\begin{figure}[tbh]
\centering
\includegraphics[width=8cm]{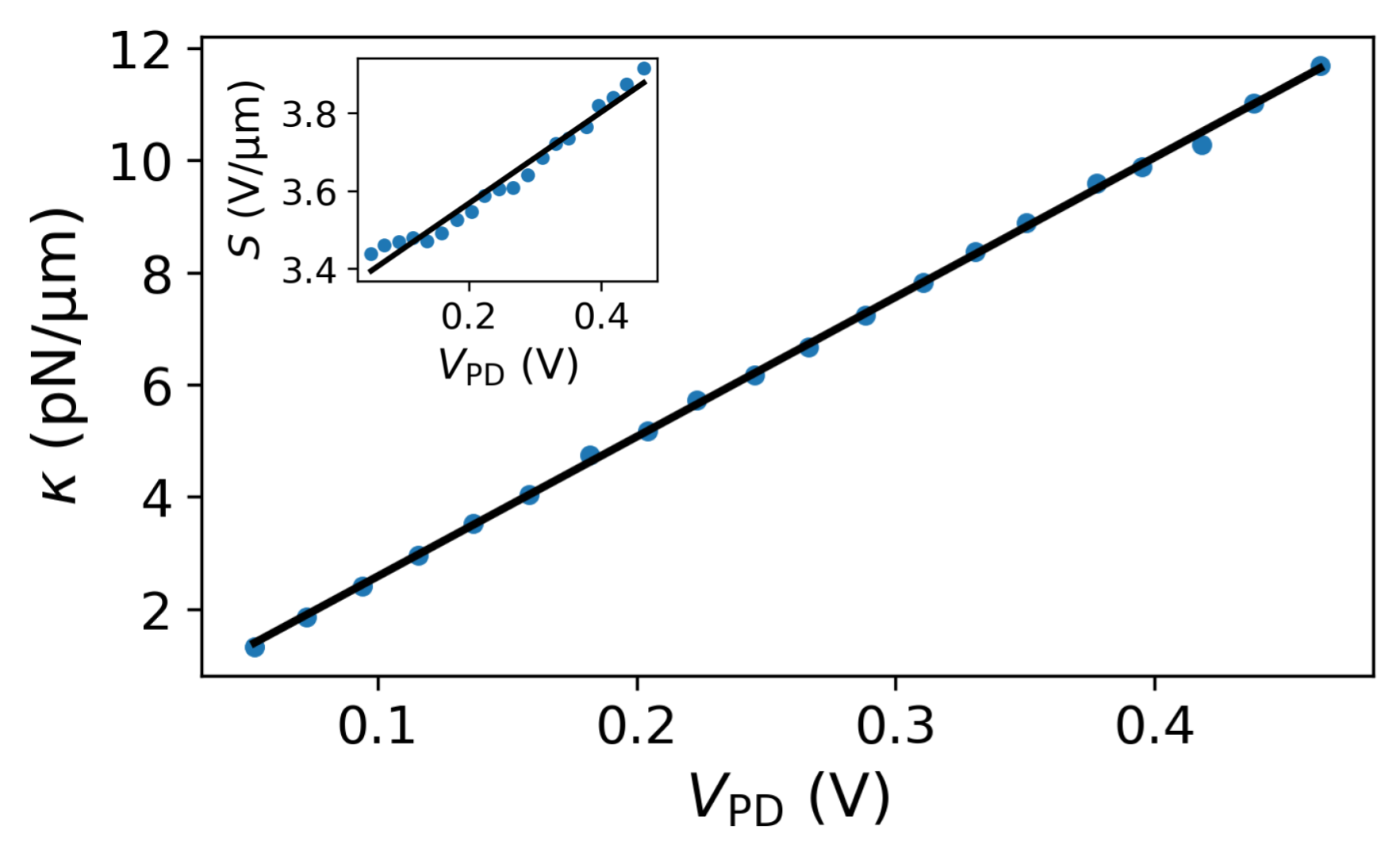}
\caption{Calibration curves for trap stiffness and amplification factor obtained from PSD fits as a function of the infrared laser power, monitored by $V_{PD}$. The trap stiffness calibration yields a slope $a_{\mathrm{TS}} = 24.9 \pm 0.1 \; \mathrm{(pN/\mu m)/V}$ and intercept $b_{\mathrm{TS}} = 0.09 \pm 0.04 \; \mathrm{pN/\mu m}$. The amplification factor calibration (inset) provides $a_{S} = 1.17 \pm 0.05 \; \mathrm{\mu m^{-1}}$, and $b_{S} = 3.33 \pm 0.01 \; \mathrm{V/\mu m}$.}
\label{fig:TS_cal_Lyon}
\end{figure}

Figure \ref{fig:TS_cal_Lyon} presents the calibration curves for the trap stiffness and amplification factor obtained from the PSD data. A linear relationship between trap stiffness and laser power is observed as expected. Notice that the amplification factor also varies slightly ($\approx12\%$) with laser power due to radiation pressure changing the equilibrium position along the $z$-axis, which affects $S_{x(y)}$. Using the data, we establish a calibration curve for the QPD response:
\begin{equation}
    S_{\mathrm{fit},x} = a_{S} V_{\mathrm{PD}} + b_{S},
\end{equation}
where $V_{\mathrm{PD}}$ is the infrared PD signal, and $a_{S}$ and $b_{S}$ are the slope and intercept of the linear fit (inset of Fig.~\ref{fig:TS_cal_Lyon}).

By simultaneously recording the QPD ($x_{\mathrm{QPD}}$) and infrared PD signals, we determine the particle's center of mass position:
\begin{equation}
    x(t) = \frac{x_{\mathrm{QPD}}(t)}{S_{\mathrm{fit},x}},
\end{equation}
and estimate the trap stiffness:
\begin{equation}
    \kappa = a_{\mathrm{TS}} V_{\mathrm{PD}} + b_{\mathrm{TS}},
\end{equation}
where $a_{\mathrm{TS}}$ and $b_{\mathrm{TS}}$ are the slope and intercept of the linear fit, as shown in Fig.~\ref{fig:TS_cal_Lyon}.

\subsection{Results and analysis}

\subsubsection{Protocol application \& work inference}

\begin{figure}[tbh]
\centering
\includegraphics[width=7cm]{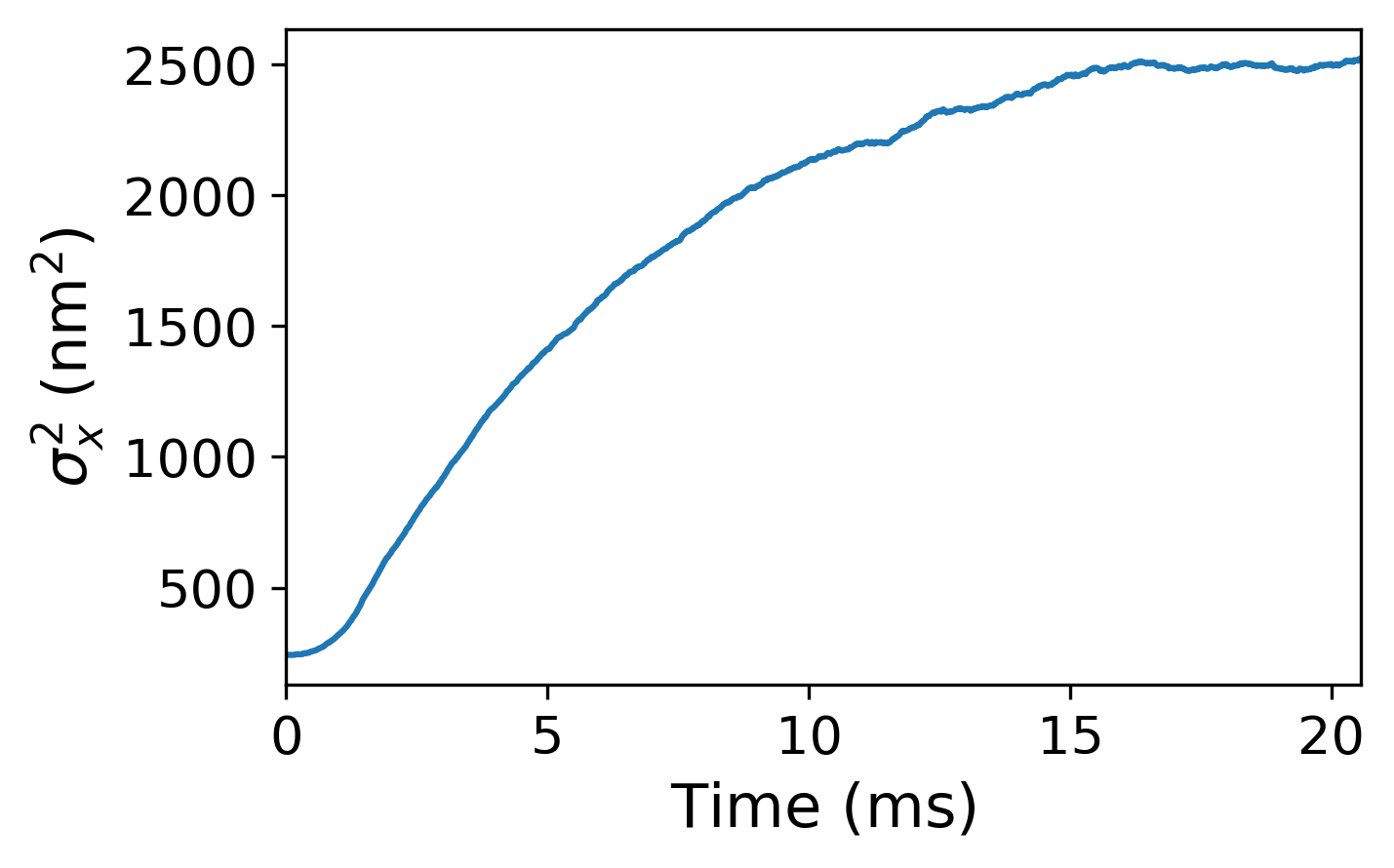}
\vspace{-3mm}
\caption{Position variance during thermalization at $\lambda_i =1.5 \ \mathrm{pN/\mu m}$, starting from $11 \lambda_i$. A linear protocol with duration $\tau_P=0.1 \tau_R$ is applied, followed by thermalization with equilibration time $\tau_{eq}=20 \ \mathrm{ms}$. This case was selected because it is the weakest trap stiffness used in the experiments here, with the longest equilibration time.}
\label{fig:variance}
\end{figure}

After characterizing the optical potential, the protocols for \textit{compression} and \textit{expansion} (see Section \ref{BreathingParabola}) were applied to the same particles used in the calibrations, using the linear protocol in Eq.~\eqref{eq:linearprotU}. To ensure thermal equilibrium, the laser power is kept constant in the initial state $\lambda_i$ for a time interval $\tau_{eq} > \tau_R$, where $\tau_R = \gamma/\lambda_i$ is the relaxation time in the state $\lambda_i$. Fig.~\ref{fig:variance} shows the measured position variance of the particle, $\sigma_x^2$, during the equilibration period at $\lambda_i = 1.5 \ \mathrm{pN/\mu m}$, demonstrating that $\tau_{eq}=20 \ \mathrm{ms}$ is sufficient for the variance to stabilize and reach a steady state even for the weakest trap.

After equilibration, the protocol is executed by linearly increasing the trap stiffness until it reaches a final value $\lambda_f > \lambda_i$. The laser intensity is then maintained until the particle thermalizes again at $\lambda_f$. This constitutes the \textit{compression} process. To execute the \textit{expansion} protocol, after the equilibration period at $\lambda_f$, the particle is linearly returned to the initial value $\lambda_i$, and the process is repeated thousands of times. 

To calculate the work during the protocol execution, one can discretize Eq.~\eqref{eq:workused}, considering that data is acquired at regular time intervals $\Delta t = 1/f_s$, leading to the cumulative work at instant $t_n$ given by \footnote{We do not use the derivative $d \lambda / dt = \Delta \lambda /\tau_P$; instead, we compute it directly from light intensity measurements in the PD output signal.}
\begin{equation}
W_{n} = \frac{1}{2} \Delta t \sum_{j=0}^{n-1} \frac{\lambda_{j+1} - \lambda_j}{\Delta t} x_j^{2} = \frac{1}{2} \sum_{j=0}^{n-1} (\lambda_{j+1} - \lambda_j) x_j^{2},
    \label{eq:workfinal}
\end{equation}
where $t_j = j \Delta t$ with $j = 0, 1, 2, \dots, n$. For a protocol with duration $\tau_P$, the temporal evolution of the work can be expressed in terms of the time step $n$. Therefore, the final work value is calculated using $n = \tau_P / \Delta t$. It is important to note that work only contributes when $d \lambda / dt \neq 0$; which means that the thermalization period does not contribute to work.

\begin{figure}[tbh]
\centering
\includegraphics[width=8cm]{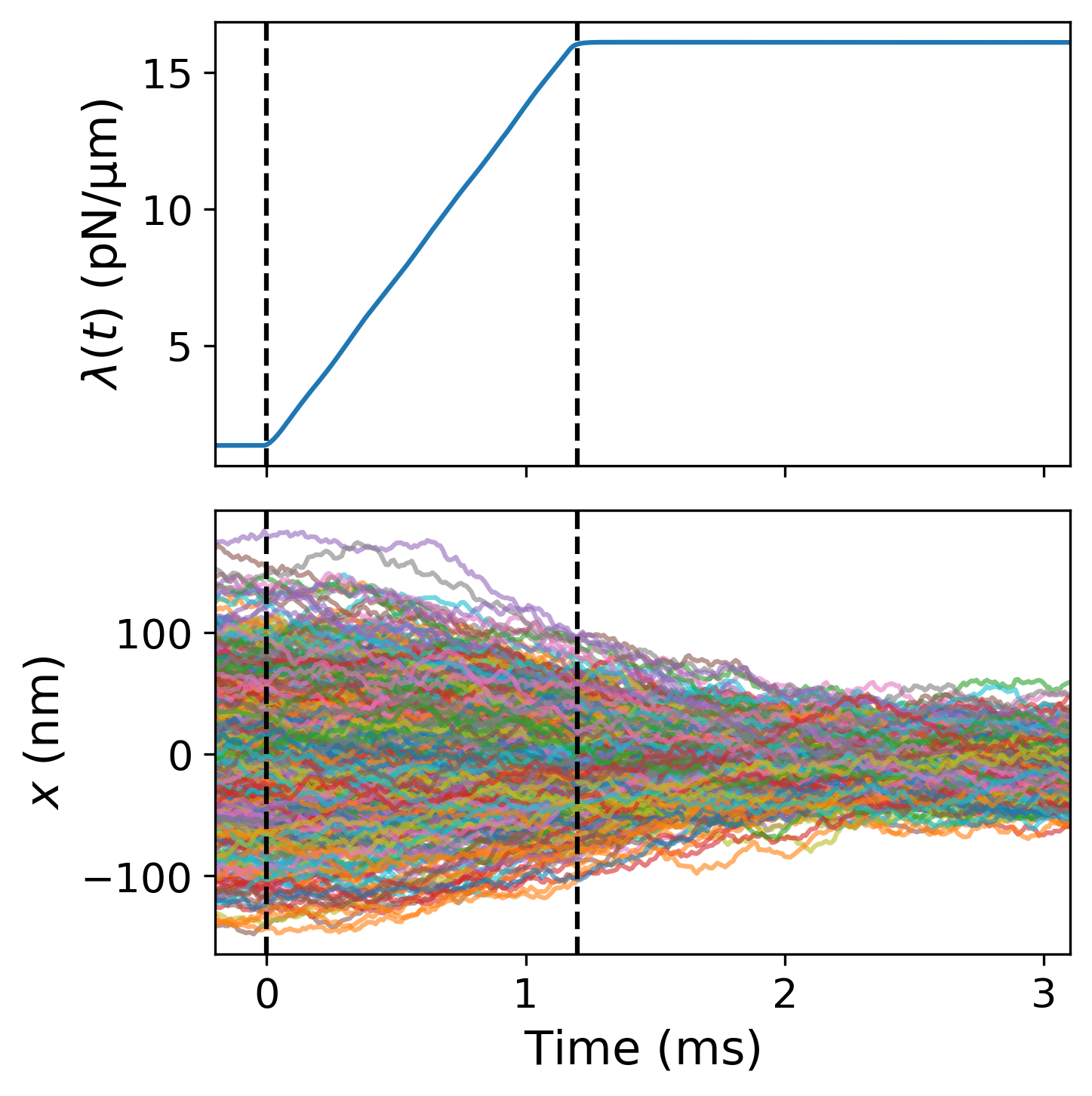}
\vspace{-2mm}
\caption{Measured control parameter (top) and 500 stochastic trajectories (bottom) for linear compression protocol with $\tau_{eq}=20 \ \mathrm{ms}$, $\lambda_i=1.5 \ \mathrm{pN/\mu m}$, $\Delta \lambda=10 \ \lambda_i$, and $\tau_P=0.1 \tau_R$, where $\tau_R = \gamma/\lambda_i \approx 12 \ \mathrm{ms}$ is the relaxation time in the state $\lambda_i$. The black dashed lines indicate the beginning and end of the protocol.}
\label{fig:cycles}
\end{figure}

After applying the protocol and recording the particle's one-dimensional position, $x_j$, simultaneously with the stiffness of the trap, $\lambda_j$, one has all the information needed to compute the relevant thermodynamic quantities.
To illustrate the experimental results of this step, Figure \ref{fig:cycles} shows the stochastic position measurements for 500 repetitions of the compression protocol, with $\tau_P=0.1 \tau_R$ and $\Delta \lambda =10 \lambda_i$.

\begin{figure}[tbh]
\centering
\includegraphics[width=7.6cm]{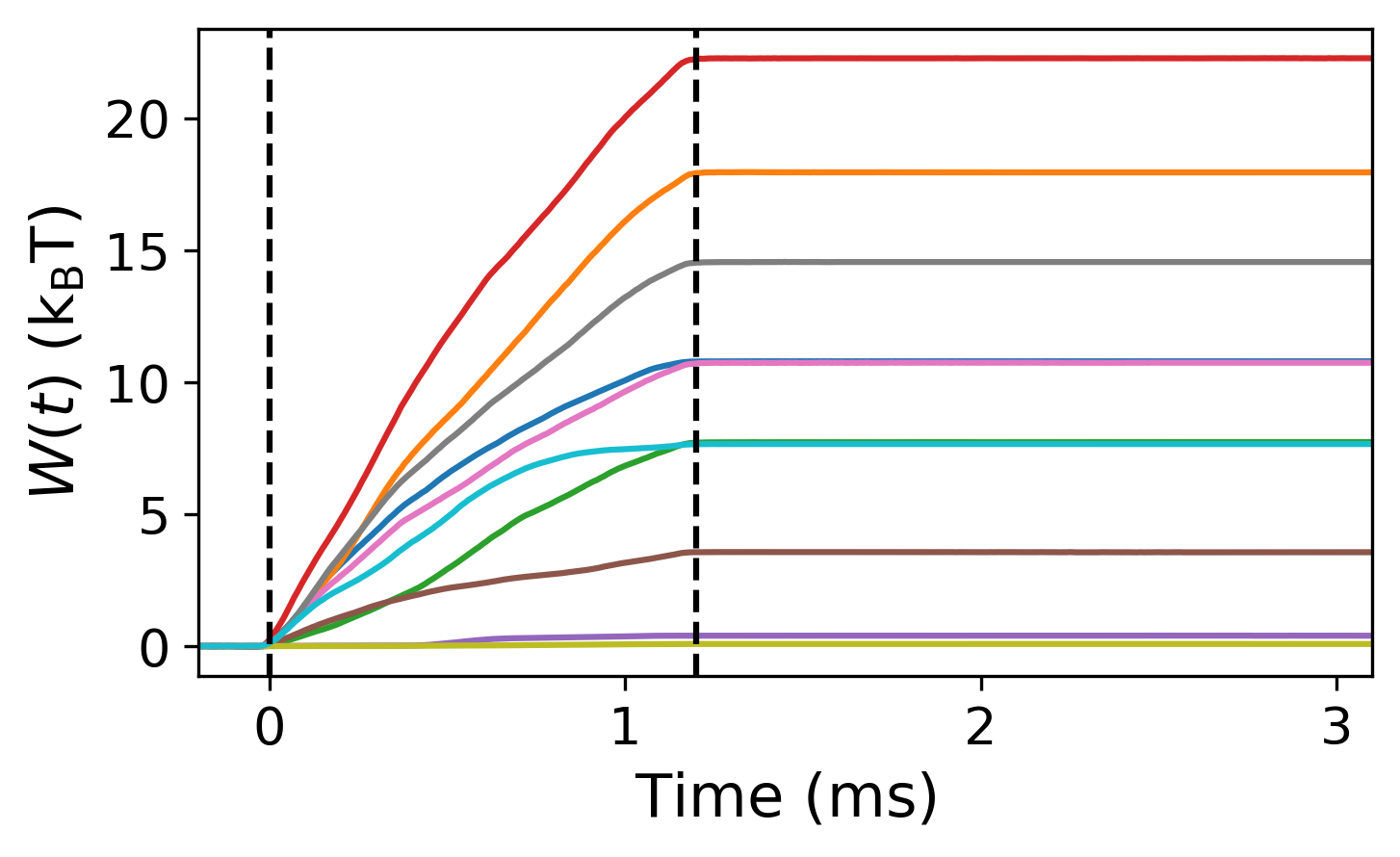}
\vspace{-2mm}
\caption{Temporal evolution of work for 10 arbitrary trajectories using the same linear compression protocol, with $\lambda_i=1.5 \ \mathrm{pN/\mu m}$, $\Delta \lambda=10 \ \lambda_i$, and $\tau_P=0.1 \tau_R$, where $\tau_R \approx 12 \ \mathrm{ms}$, and $\tau_{eq}=20 \ \mathrm{ms}$. The black dashed lines indicate the beginning and end of the protocol.}
\label{fig:Workloop}
\end{figure}

Using Eq.~\eqref{eq:workfinal} and the acquired data, one calculates the fluctuating work across the stochastic trajectories. Figure \ref{fig:Workloop} displays the results for 10 distinct runs of the same protocol.

\begin{figure}[tbh]
\centering
\begin{tabular}{ll}
\includegraphics[width=7.8cm]{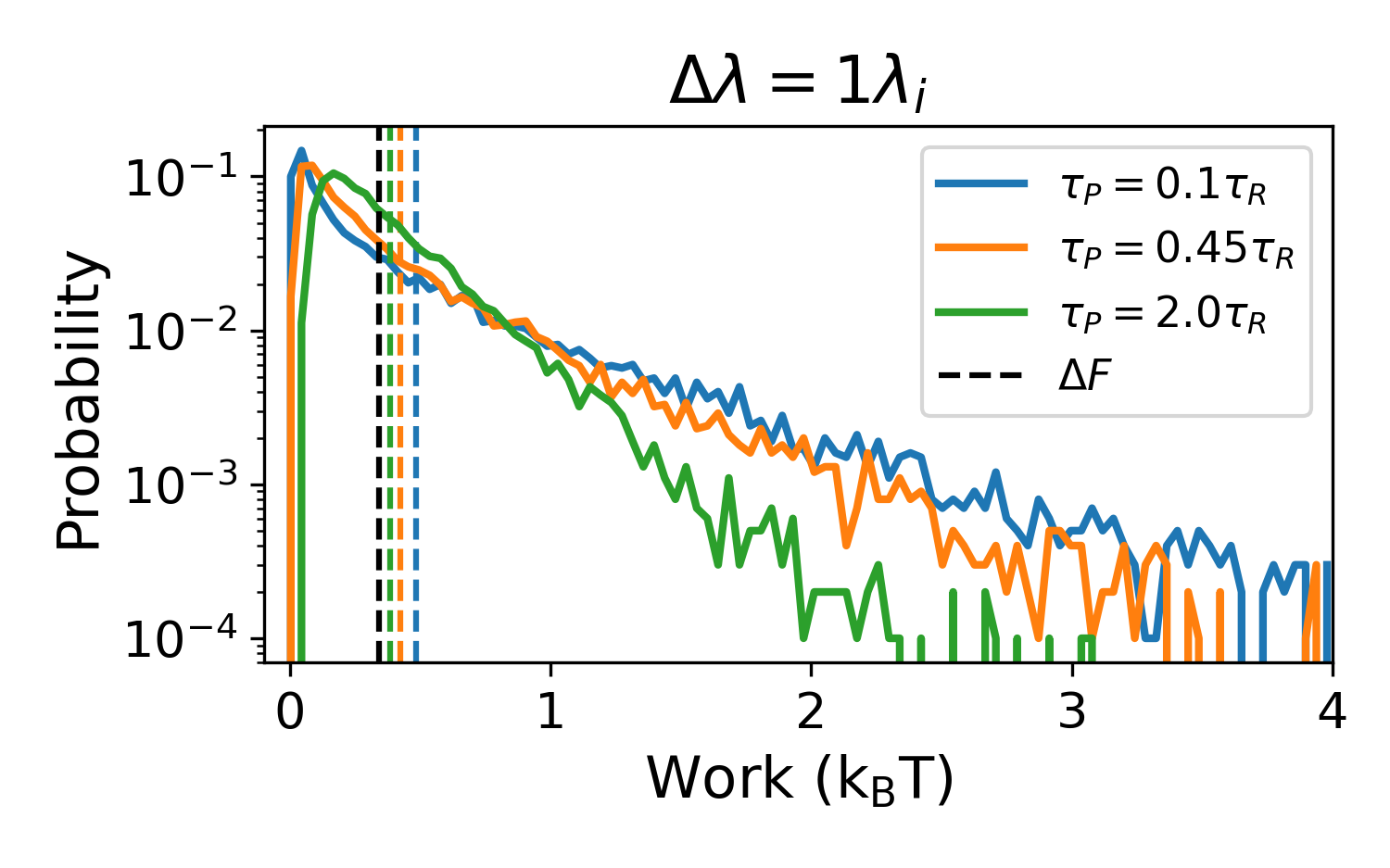}
\\
\includegraphics[width=7.8cm]{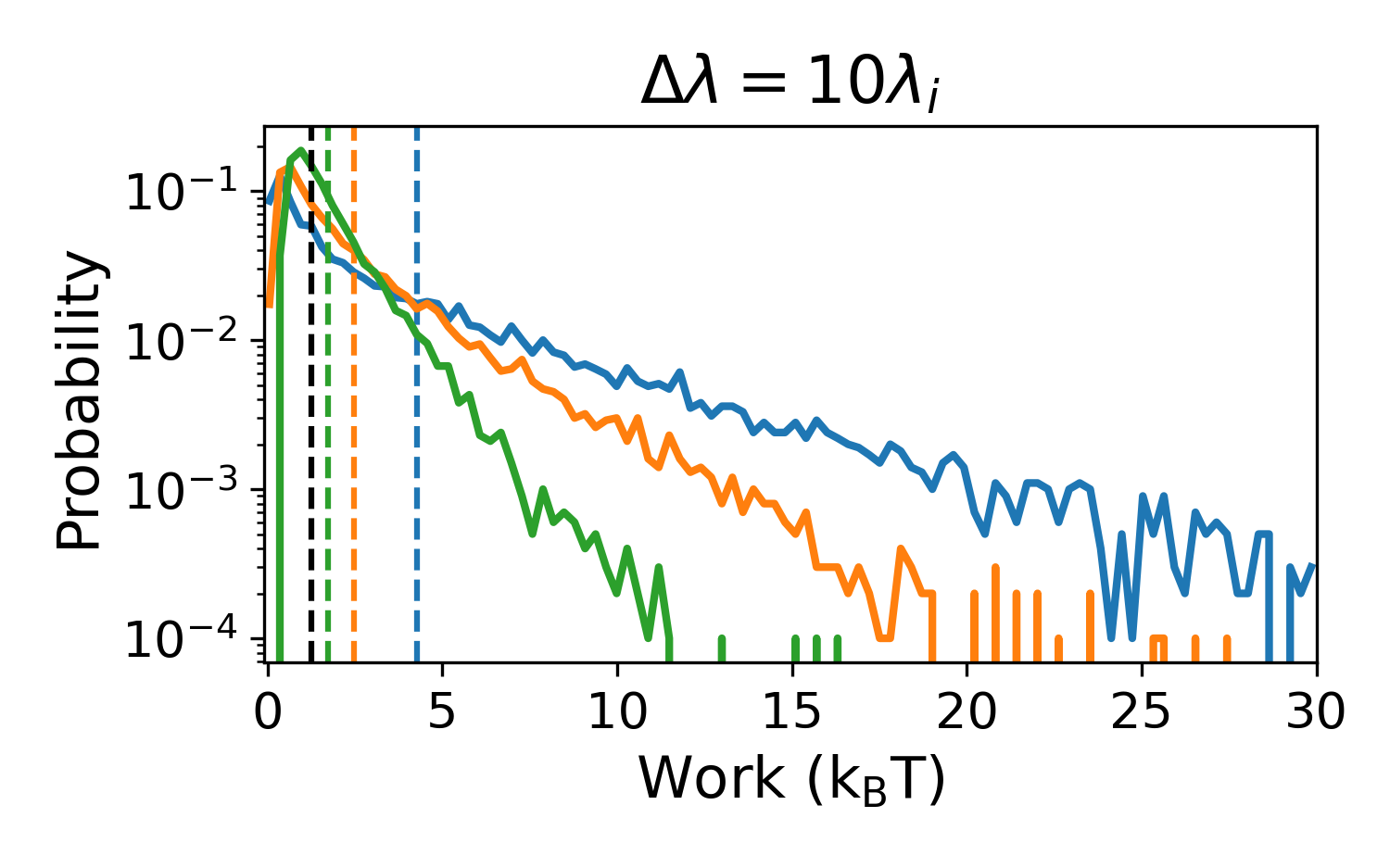}
\end{tabular}
\vspace{-2mm}
\caption{Work probability distributions (colored solid lines) and the respective average value of work (same color dashed lines) for protocol times: $\tau_P = \{0.1\tau_R,\, 0.45\tau_R,\, 2\tau_R \}$, in blue, orange, and green respectively. Two modulation amplitudes are considered, $\Delta \lambda= \lambda_i$ (top) and $\Delta \lambda=10 \ \lambda_i$ (bottom). Results are from $N=10^4$ trajectories of forward linear protocols with $\tau_{eq}=20 \ \mathrm{ms}$, $\lambda_i=1.5 \ \mathrm{pN/\mu m}$, and $\tau_R \approx 12 \ \mathrm{ms}$. The black dashed line indicates the difference in free energy. The number of bins used for the histograms is 100.}
\label{fig:Histo_timef}
\end{figure}

After a large number of repetitions ($N=10^4$) of the same protocol, a histogram is produced using the calculated work and the probability density function (pdf), $P(W)$, is determined, ensuring $\int P(W) \, dW = 1$. Figure~\ref{fig:Histo_timef} shows the results for two modulation amplitudes ($\Delta \lambda$) and three different protocol durations ($\tau_{P}$). For each colored solid curve, a vertical dashed line indicates the average work, while the free energy change, $\Delta F$, is marked by the black vertical dashed lines. As expected, the measured work exhibits significant fluctuations with $\langle W \rangle > \Delta F$. Fig.~\ref{fig:Histo_timef} also allows a direct comparison of the work pdf between far-from-equilibrium (quick, $\tau_P=0.1\tau_R$) and near-equilibrium (slower, $\tau_P=2\tau_R$) protocols.

Examining these distributions, we see that the work variance decreases and the average work approaches the free-energy difference as the protocol is driven more slowly. This behavior is consistent with the second law, as described in Eq.~\eqref{2law2}. In the ideal quasistatic scenario (where the process is infinitely slow), one would expect a succession of equilibrium states with a narrow distribution centered at the free energy difference $\Delta F$, the minimum required to change the state. 

In particular, although the average work never falls below the free energy difference, individual trajectories can occasionally be below $\Delta F$. These instances have sometimes been called "violations" of the second law. 
The frequency of these rare events increases as the process deviates further from equilibrium, i.e., with larger modulation amplitudes and shorter protocol durations. Nevertheless, the distributions of the works always satisfy
$\int_{-\infty}^{\Delta F} P(W) \, dW < \int_{\Delta F}^{\infty} P(W) \, dW$.

\subsubsection{Jarzynski equality}

Building on previous steps, we can verify JE, Eq.~\eqref{jarzynski}. As mentioned earlier, the exact free energy difference for the linear protocol is given by Eq.~\eqref{eq:deltafeq}. Therefore, the right-hand side of Eq.~\eqref{jarzynski} can be expressed simply as 
\begin{equation}
e^{-\Delta F / k_B T} = \frac{Z_f}{Z_i} = \sqrt{\frac{\lambda_i}{\lambda_f}},
\end{equation}
where $Z_{\{i,f\}}$ and $\lambda_{\{i,f\}}$ are the partition functions and the trapping control parameters of the initial and final states.

Using the work statistics obtained from the experimental data, the left-hand side of Eq.~\eqref{jarzynski} can also be calculated.

The uncertainty of \(e^{-\Delta F / k_B T}\) is calculated using uncertainty propagation:
\begin{equation}
    \delta \left[e^{-\Delta F / k_B T}\right] = \frac{1}{2} \sqrt{\left(\frac{\delta \lambda_i}{\lambda_f}\right)^2 + \left(\frac{\lambda_i \delta \lambda_f}{\lambda_f^2}\right)^2},
\end{equation}
where \(\delta \lambda(t) = \sqrt{(\delta m_{TS} V_{PD}(t))^2 + (\delta V_{PD}(t) m_{TS})^2 + (\delta b_{TS})^2}\) represents the uncertainties in the stiffness of the trap in the initial (\(\delta \lambda_i\)) and final (\(\delta \lambda_f\)) states. The uncertainty of \(\delta \left[ \langle \exp(-W / k_B T) \rangle \right]\) is determined by dividing the dataset into subsets (blocks) and using the block averaging method \cite{Allen-Tildesley2017, Jonsson2018pre, Flyvbjerg1989}. Specifically, we computed the averages \( \mu_j = \langle \{\exp(-W_n / k_B T)\}_{n=1,\ldots,1000} \rangle \) for 10 sets ($j=1,\ldots,10$), each set (block) containing 1000 trajectories. The uncertainty \(\delta \left[ \langle \exp(-W / k_B T) \rangle \right]\) was then estimated from the standard deviation of these values (\(\sigma_{\langle \mu_j \rangle}\)), calculated as the standard error of the mean for these independent sets: \(\sigma_{\langle \mu_j \rangle} / \sqrt{10}\). 

\vspace{-1mm}
\begin{figure}[tbh]
\centering
\includegraphics[width=7.9cm]{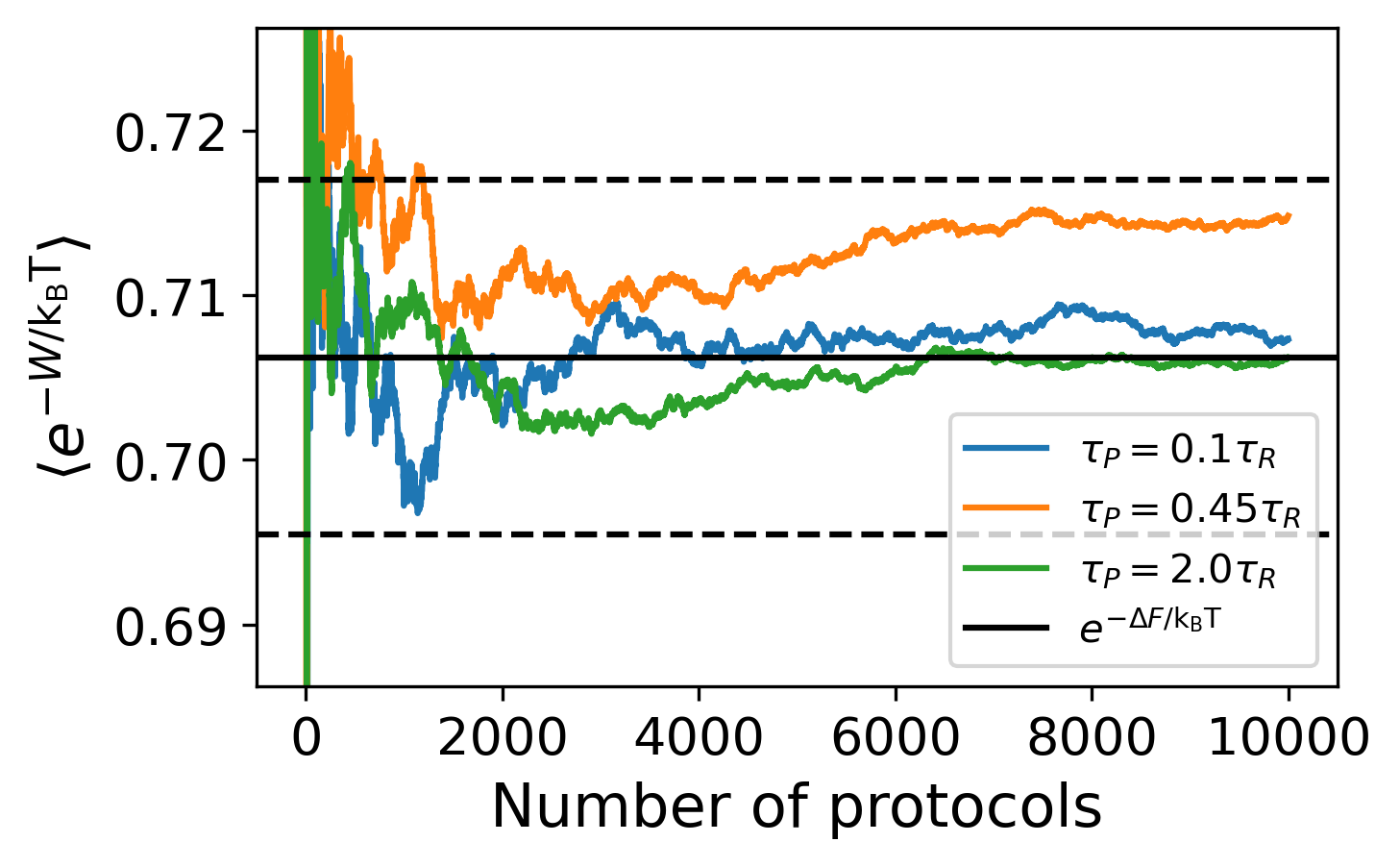}
\vspace{-3mm}
\caption{Convergence of \(\langle e^{-W / k_B T} \rangle\) versus the number of experimental executions of the linear compression protocol for \(\Delta \lambda = \lambda_i\) and same parameters of Fig.~\ref{fig:Histo_timef}. The black solid center line corresponds to \(e^{-\Delta F / k_B T}\), and dashed lines represent the uncertainty range.}
\label{fig:ConvJarzynski}
\end{figure}

Figure \ref{fig:ConvJarzynski} illustrates the behavior of \(\langle e^{-W / k_B T} \rangle\) as a function of the number of repeats of the protocol. Note that the average value converges towards a stable value as the number of trajectories increases. Ideally, an infinite number of samples would be required for an exact verification, but the results show that $N\approx2000$ realizations are sufficient to confirm the Jarzynski equality within the calculated uncertainties.

\begin{figure}[tbh]
\centering
\includegraphics[width=7.6cm]{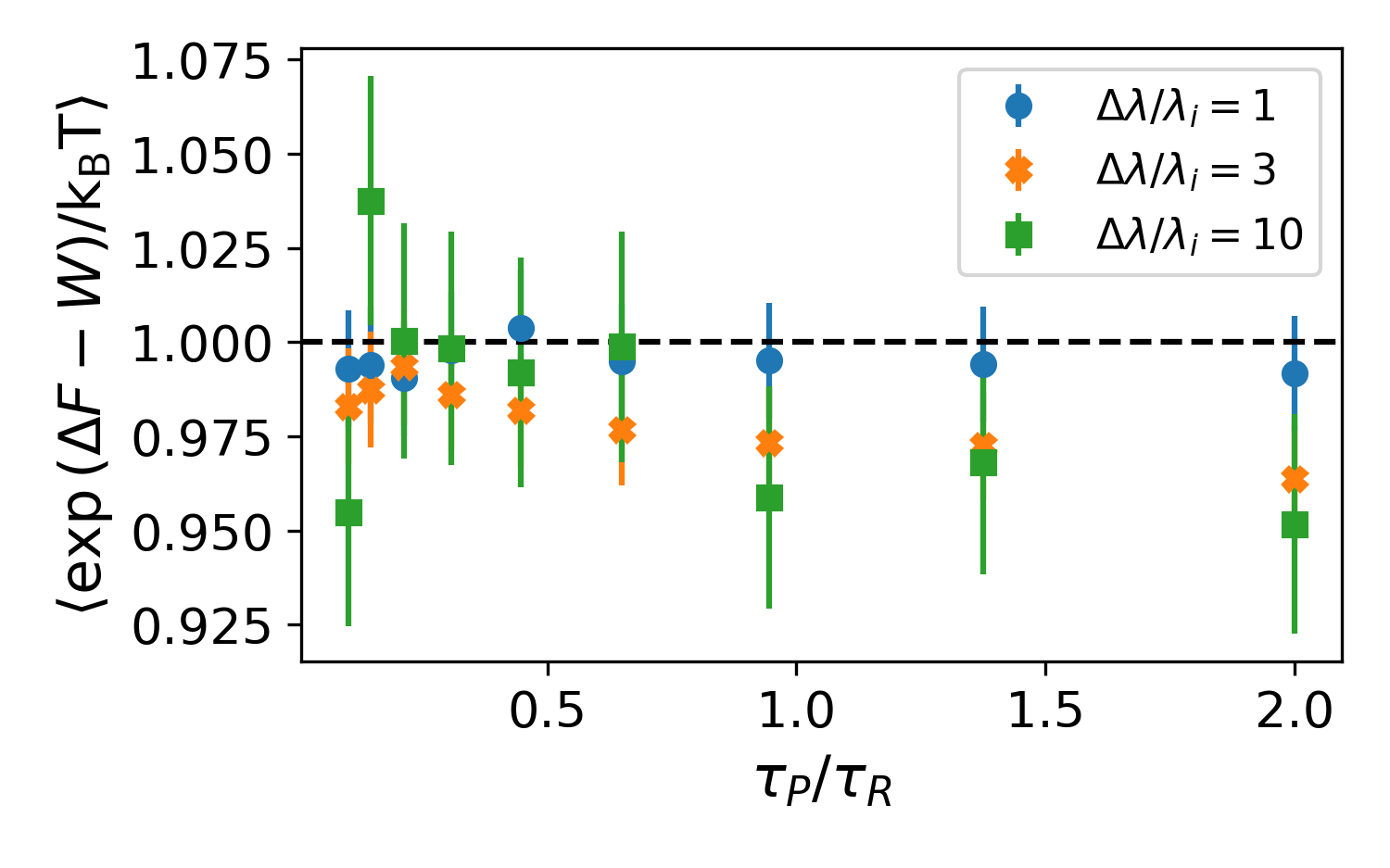}
\vspace{-2mm}
\caption{Verification of Jarzynski equality for a linear compression. The average $\langle e^{(\Delta F-W)/ k_B T} \rangle$ for different durations $\tau_P$ and modulation amplitudes: $\Delta \lambda / \lambda_i = \{1,\ 3,\ 10\}$, in blue, orange, and green, respectively. Each point represents data from $10^4$ realizations of the experiment at the same conditions. The black dashed line shows the (exact) expected value, and the bars indicate the uncertainty for each point.}
\label{fig:WLJarzynski}
\end{figure}

Figure \ref{fig:WLJarzynski} presents the verification of the Jarzynski equality across different durations of the protocol (from \(\tau_P = 0.1\tau_R\) to \(\tau_P = 2\tau_R\)) and modulation amplitudes ($\Delta \lambda = \{\lambda_i,\,3 \lambda_i,\, 10 \lambda_i\}$), all with \(N=10^4\) realizations. The results indicate that for small amplitude modulation (\(\Delta \lambda = \lambda_i\)), the JE holds consistently across all examined protocol durations. More specifically, the value of \(e^{-\Delta F / k_B T}\) obtained from the PD readings during the initial and final equilibrium periods aligns well with \(\langle e^{-W / k_B T} \rangle\) obtained from the work distributions. For higher values of $\Delta \lambda$, a better agreement is observed with shorter protocol times. This discrepancy is likely due to the miscalibration of the amplification factor \(S_x\) depending on the power of the trapping beam, as we explain later.

\begin{figure}[tbh]
\centering
\includegraphics[width=7.8cm]{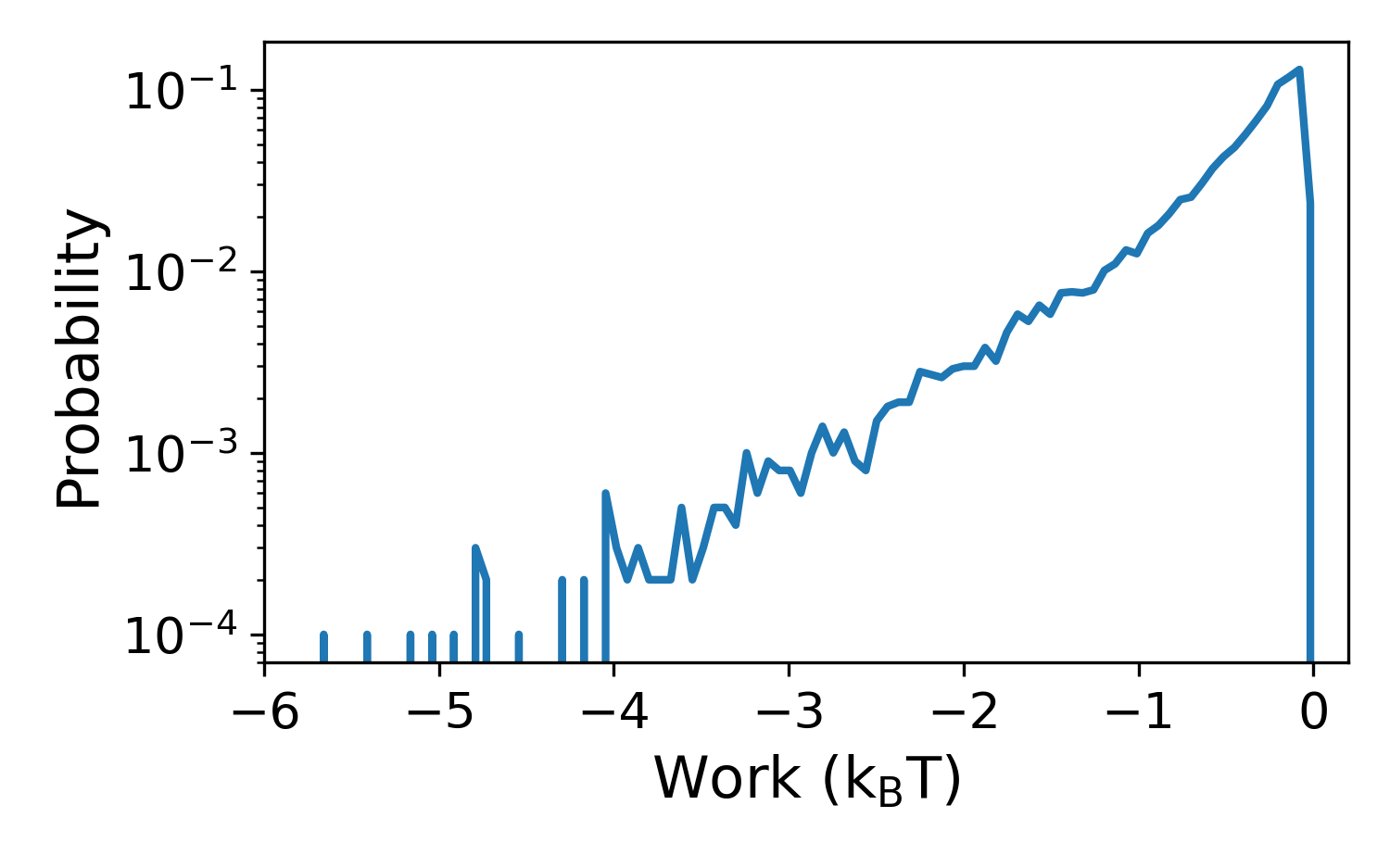}
\vspace{-2mm}
\caption{Probability distributions of work for the linear expansion protocol with \(N=10^4\) trajectories, \(\tau_{eq} = 20 \ \mathrm{ms}\), \(\lambda_i = 1.5 \ \mathrm{pN/\mu m}\), \(\tau_P = 0.1 \ \tau_R\), \(\tau_R \approx 12 \ \mathrm{ms}\), \(\Delta \lambda = 10 \ \lambda_i\), and using 100 bins.}
\label{fig:Histo_fr_v3_10}
\end{figure}

\begin{figure}[tbh]
\centering
\includegraphics[width=8cm]{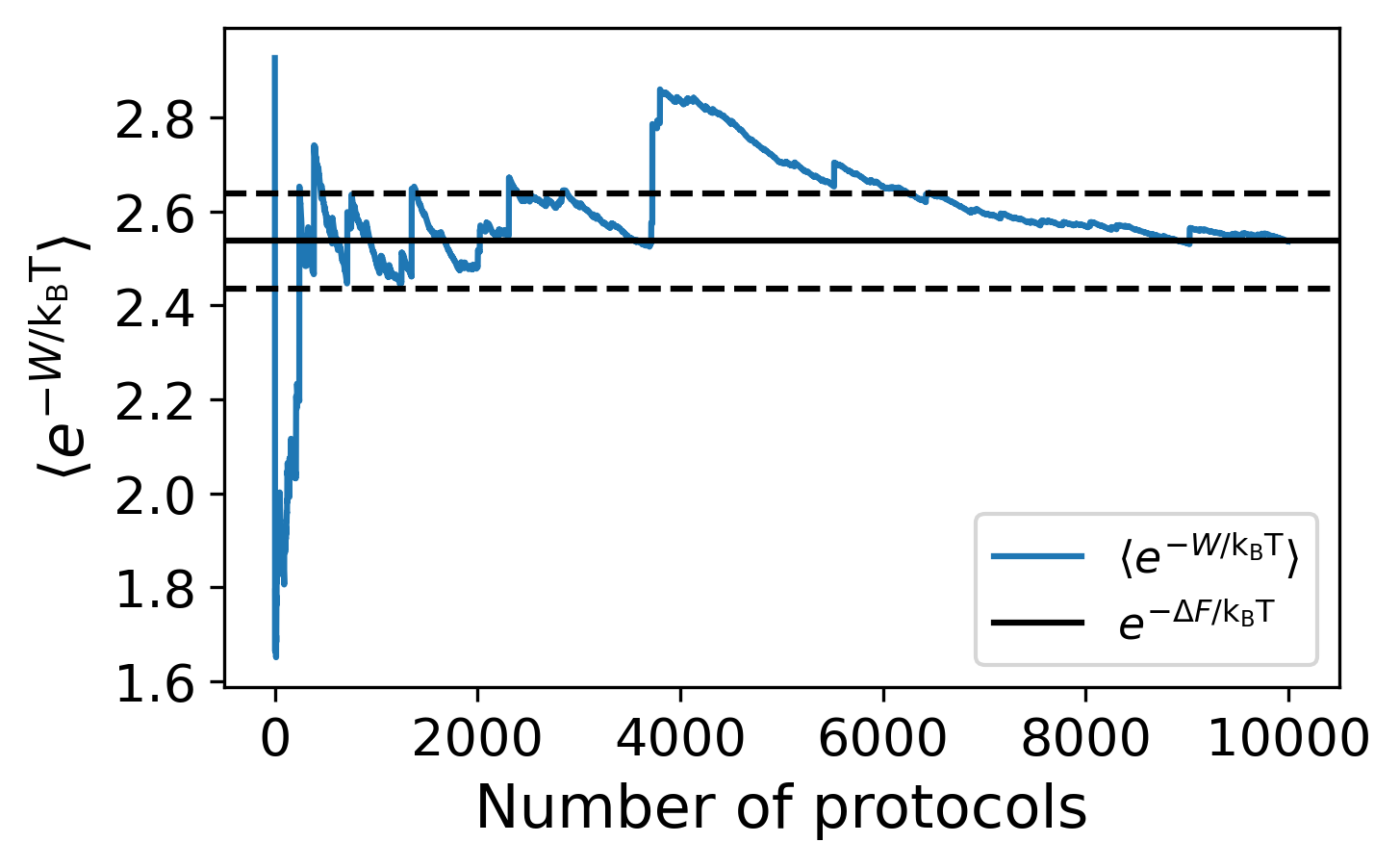}
\vspace{-2mm}
\caption{Convergence of \(\langle e^{-W / k_B T} \rangle\) versus the number of experimental executions of the linear expansion protocol for \(\Delta \lambda = 10 \ \lambda_i\), \(\tau_P = 0.21 \ \tau_R\), \(\tau_{eq} = 20 \ \mathrm{ms}\), \(\lambda_i = 1.5 \ \mathrm{pN/\mu m}\), and \(\tau_R \approx 12 \ \mathrm{ms}\). The black dashed lines represent the uncertainty range. Notice the significant effect of rare events (sudden jumps) in the convergence.}
\label{fig:ConvJarzynskireverse}
\end{figure}

Interestingly, in the linear expansion protocol, the stiffness of the trap decreases over time, i.e., \(d\lambda/dt < 0\), leading to negative work values since \(\langle x^2 \rangle > 0\), as shown in Fig.~\ref{fig:Histo_fr_v3_10}. It is important to understand that here the rare events (occurring in the tails of the probability distributions) significantly impact the calculation of \(\langle e^{-W / k_B T} \rangle\). This illustrates a general caveat when using the Jarzynski estimator in regimes where rare events dominate convergence: although the equality holds exactly, convergence can be prohibitively slow, as shown in Fig.~\ref{fig:ConvJarzynskireverse}. Most repetitions have little influence on the average value, while a few rare events greatly affect the exponential average \cite{jarzynski2006rare,Kamizaki2022}. In cases like this, it is better to use the Crooks relation, Eq.~\eqref{eq:Crooks}, because it uses the probabilities ratio and is less affected by these rare events.

\subsubsection{Crooks fluctuation theorem}

As introduced in Section \ref{FlucTheorems}, Gavin Crooks proposed Eq.~\eqref{eq:Crooks}, which relates the probabilities of work in a forward protocol ($P_{F}(W)$) and its backward protocol (time-reversed) ($P_{R}(-W)$). In our experiment, the expansion (reverse) protocol was applied after the compression (forward) protocol, allowing sufficient time for equilibration before proceeding, as explained previously.

\begin{figure}[tbh]
\centering
\begin{tabular}{ll}
\includegraphics[width=8cm]{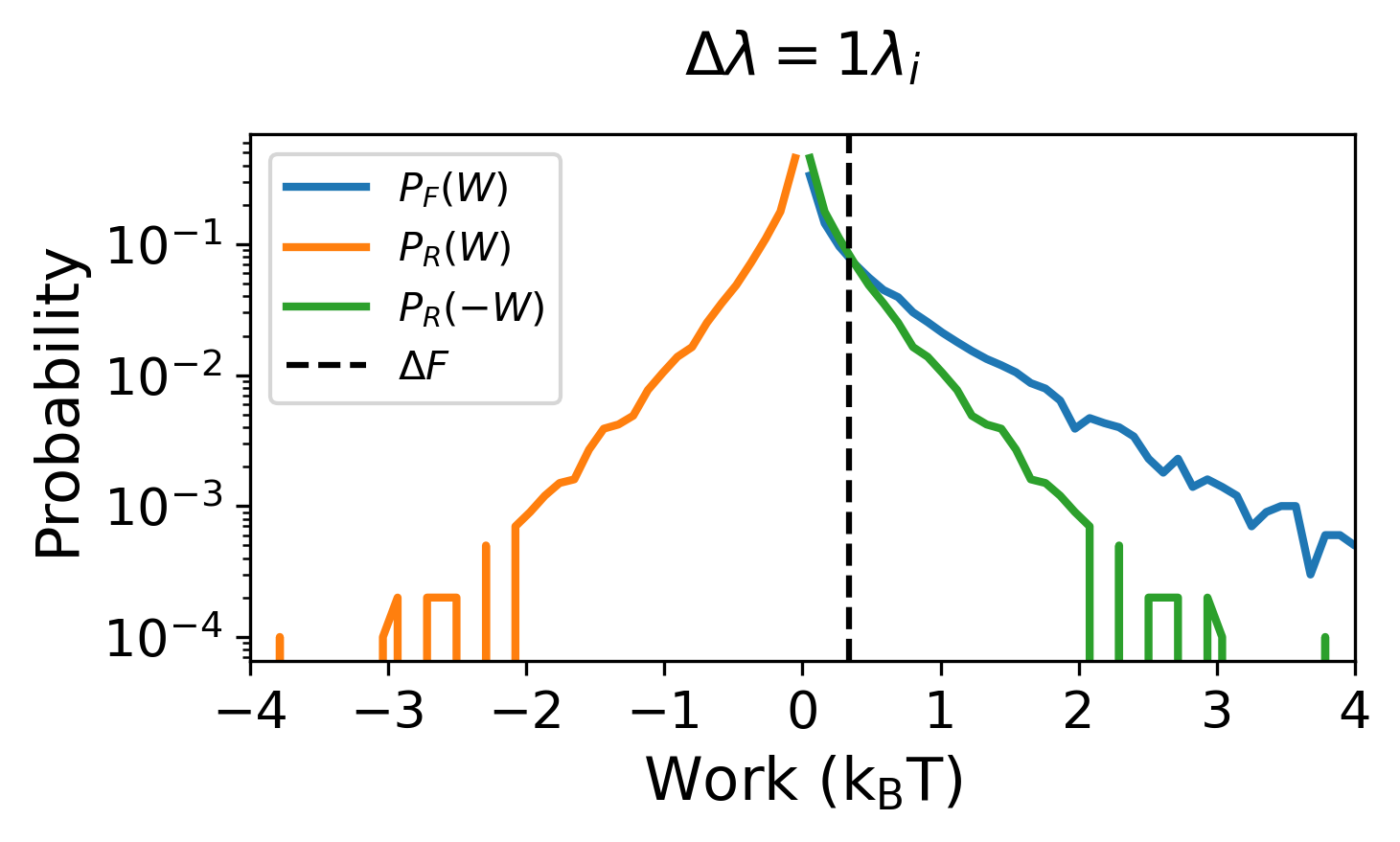}
\\
\includegraphics[width=8cm]{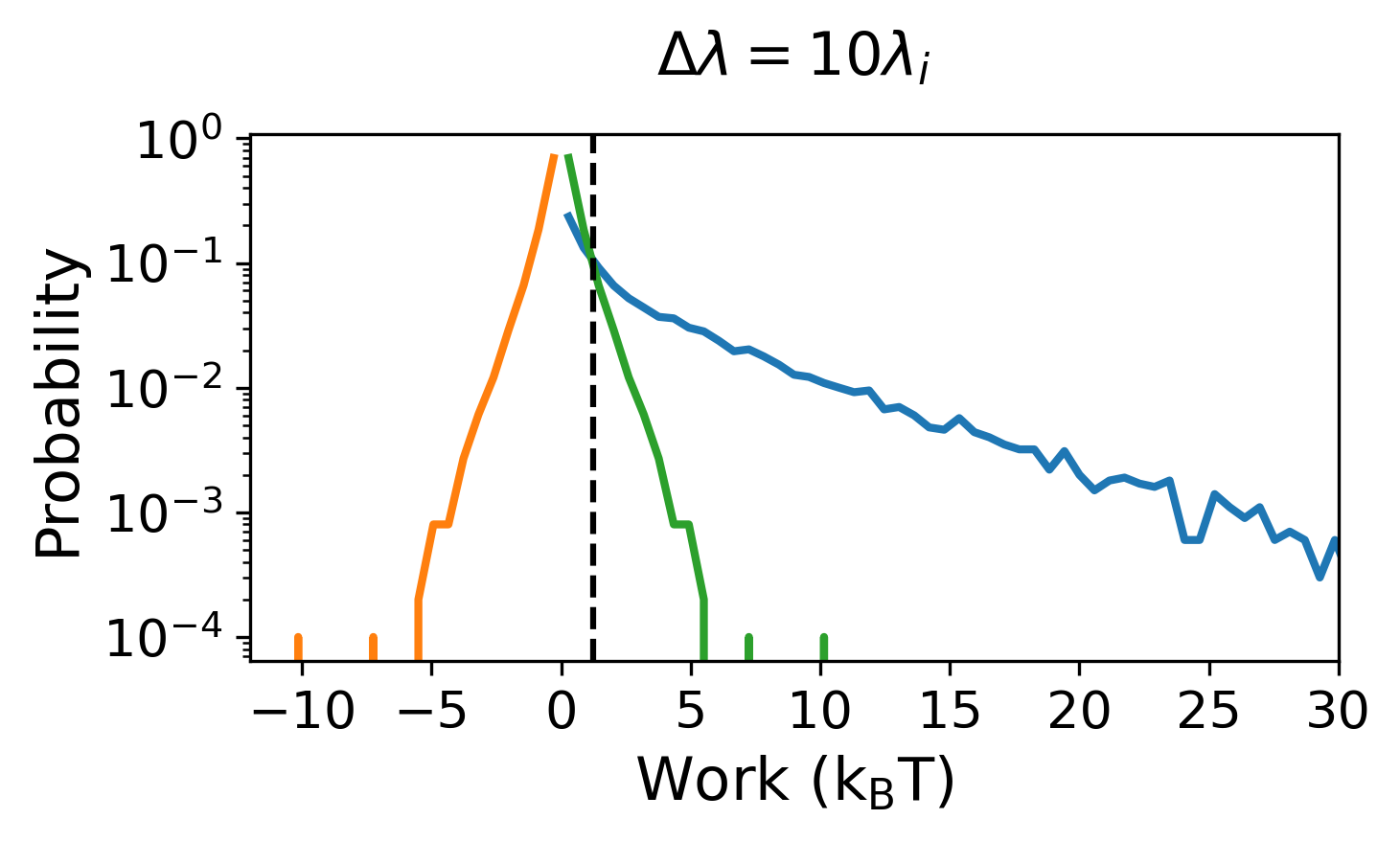}
\end{tabular}
\vspace{-1mm}
\caption{Probability distributions \(P_F(W)\), \(P_R(W)\), and \(P_R(-W)\) for $10^4$ trajectories with \(\tau_{eq} = 20\ \mathrm{ms}\), \(\lambda_i = 1.5\ \mathrm{pN/\mu m}\), \(\tau_P = 0.1\tau_R\), with \(\tau_R = \gamma/\lambda_i \approx 12\ \mathrm{ms}\), and \(\Delta \lambda = \lambda_i\) (top) and \(\Delta \lambda = 10 \lambda_i\) (bottom). The black dashed line indicates the free energy difference $\Delta F$. The number of bins is set to 100 for each range from zero to the maximum positive work value observed in the forward protocol.}
\label{fig:Histo_fr}
\end{figure}

Figure \ref{fig:Histo_fr} displays the stochastic work distributions from $N=10^4$ experimental realizations for two modulation amplitudes. As anticipated, the crossing point \(P_{F}(W) = P_{R}(-W)\) occurs precisely at the work value equal to the difference in free energy \(\Delta F\), as indicated by the vertical black dashed lines calculated using Eq.~(\ref{eq:deltafeq}). This observation provides a straightforward method to estimate \(\Delta F\). Moreover, it is worth noting that in this particular case, verifying the result from the Crooks theorem is easier and more robust than using Jarzynski because the intersection point is positioned near the peaks of the histograms, which minimizes the effect of rare events on the results \cite{jarzynski2006rare,Kamizaki2022}.

\begin{figure}[tbh]
\centering
\begin{tabular}{ll}
\includegraphics[width=7cm]{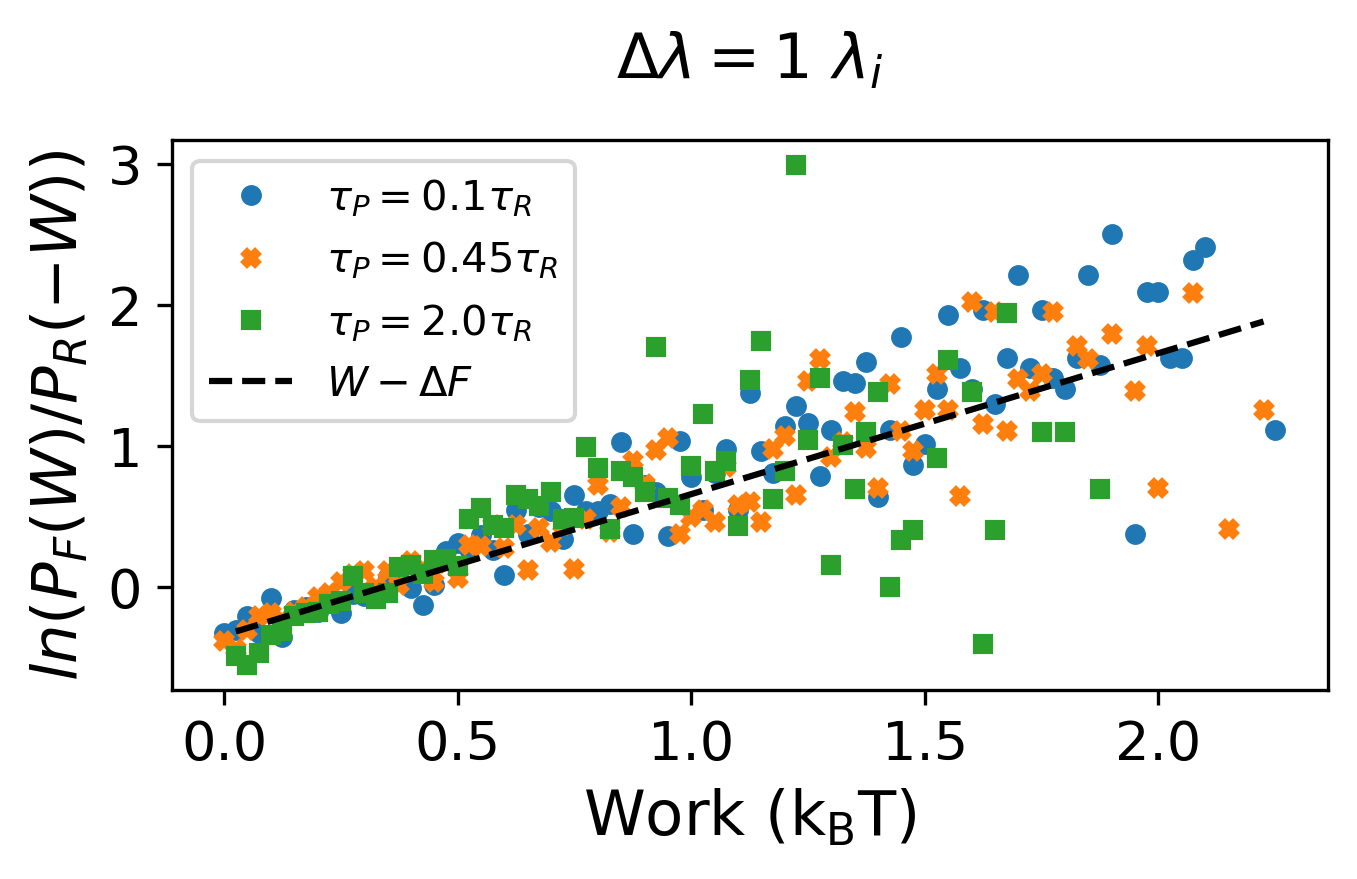}
\\
\includegraphics[width=7cm]{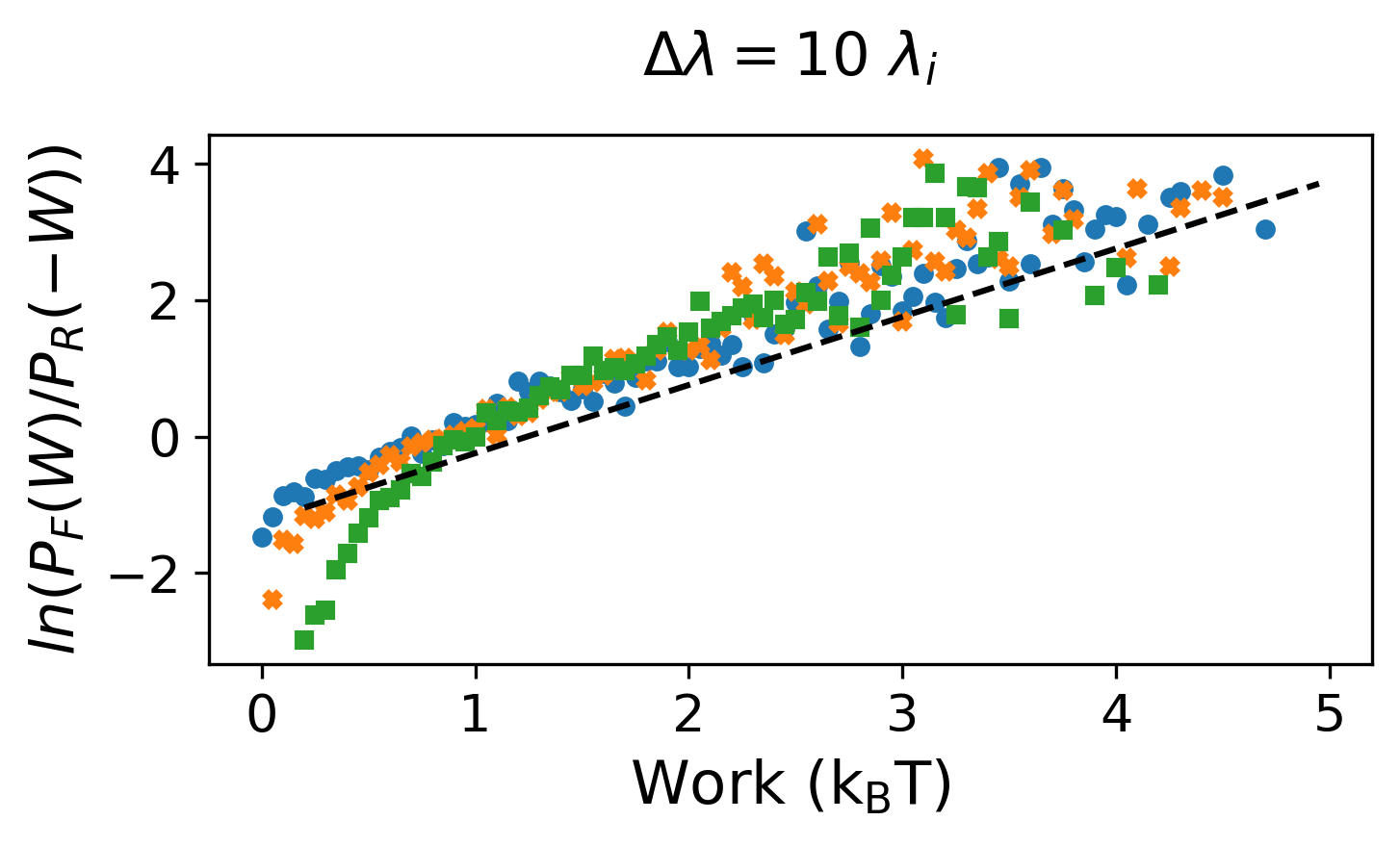}
\end{tabular}
\caption{Verification of Crooks Fluctuation Theorem for \(\Delta \lambda = \lambda_i\) (top) and \(\Delta \lambda = 10 \ \lambda_i\) (bottom). Experimental values of \(\ln\left[\frac{P_{F}(W)}{P_{R}(-W)}\right]\) for $10,000$ trajectories with \(\lambda_i = 1.5\ \mathrm{pN/\mu m}\) and different $\tau_P$.}
\label{fig:Crooks_nofit}
\vspace{-3mm}
\end{figure}

Taking the \textit{logarithm} of Eq.~\eqref{eq:Crooks} yields
\begin{equation}
    \ln\left[\frac{P_{F}(W)}{P_{R}(-W)}\right] = \frac{W - \Delta F}{k_{B} T},
    \label{eq:Crooksln}
\end{equation}
allowing us to assess the validity of the relation for various work values by plotting \(\ln\left[\frac{P_{F}(W)}{P_{R}(-W)}\right]\) versus \( W \). Figure \ref{fig:Crooks_nofit} shows the experimental values of \(\ln\left[\frac{P_{F}(W)}{P_{R}(-W)}\right]\) for different durations and amplitudes of the protocol. The data align with the linear function \( \frac{W - \Delta F}{k_{B} T} \) at lower work values with higher counts, while it scatters at larger work values due to rare events (lower counts). There is better agreement between experimental data and theoretical predictions for protocols with smaller modulation amplitudes and shorter durations, probably due to miscalibration of the amplification factor, as mentioned before. However, the data remain consistent with the expected behavior and comply with the CFT \cite{jarzynski2006rare}.

\section{Discussion and conclusions} \label{Conclusions}

This work helps to disseminate knowledge of stochastic thermodynamics by introducing and measuring the Jarzynski and Crooks relations, along with key ideas introduced by Sekimoto \cite{Sekimoto-book}. Here, we present a simple home-built apparatus and thoroughly explain the methods for measuring and validating these outstanding results of contemporary thermodynamics through experiments with optical tweezers. These non-equilibrium work relations connect an equilibrium quantity, the free energy difference, and out-of-equilibrium measurements. Notably, these relations enable the determination of free-energy differences without performing a quasistatic process, which would require prohibitively long times or would be otherwise practically impossible (e.g. in a chemical reaction) for processes for which the transformation rate cannot be directly controlled.

The experiment used finite time \textit{compression} and \textit{expansion} protocols on an optically trapped Brownian particle. By modulating the intensity of the trapping beam, we can dynamically adjust the parameters of the external potential, thereby altering the state of the particle. By characterizing the optical potential at various laser intensities and analyzing the particle's trajectory, we can compute the fluctuating work applied during each protocol execution to construct probability distributions from an ensemble of repetitions. The calculated free-energy values for the initial and final states, which are determined experimentally through the well-characterized optical potential, can be directly compared with those obtained experimentally from the work probability distributions.

Regarding Jarzynski equality, the experimental values of $\left\langle e^{- W/k_{B}T}\right\rangle$  were verified to be consistent with $e^{-\Delta F/k_{B}T}$ after a few thousand realizations of the compression protocol, being largely stable across all parameter sets. We experimented with different modulation amplitudes and protocol times to confirm the robustness of this convergence. For the expansion protocol, due to the negative work values, we showed that the Jarzynski average can be significantly affected by rare events.

Applying the Crooks relation, we showed that it is more robust and less sensitive to rare events, and we also obtained a good agreement between theory and experiments. This agreement is stronger for higher counts, with some scattering observed in the regions corresponding to rare events. 

Overall, both fluctuation theorems were verified experimentally, essentially matching theoretical predictions, with better agreement for lower modulation amplitudes and shorter protocol times. The difference at higher intensities results from a small systematic miscalibration of the amplification factor, sensitive to radiation pressure \cite{martins2024thesis}, as indicated in the inset of Fig.~\ref{fig:TS_cal_Lyon}. 
The equilibrium position of the particle at a constant intensity determines the amplification factor. When the intensity of the trapping beam varies during the protocol, the amplification factor may change as the particle moves along the $z$-axis due to radiation pressure. However, without access to the actual movement, we calculate the amplification factor based on the initial state, represented as $S_{x}=S_{fit,x}(\langle V_{PD, eq} \rangle)$, where $\langle V_{PD, eq} \rangle$ is the average reading during the equilibrium period. For fast protocols, the position of the particle along the $z$-axis does not change significantly, making this approximation sufficient.
In practice, this systematic error introduces only minor variations that do not deviate significantly from what is expected. 
These results show the versatility of optical tweezers for experiments to test and explore fundamental principles in stochastic thermodynamics.

For a long time, our understanding of thermodynamics was limited to near-equilibrium macroscopic systems. In the last few decades, adding to the toolset of statistical mechanics, the advent of stochastic thermodynamics and versatile experimental platforms like optical tweezers has provided a new set of tools to explore nonequilibrium phenomena of small systems. The true horizon for these tools, however, lies in their application to opening new frontiers and helping solve pressing energy challenges. 
In fact, the present-day drive for new, sustainable, and efficient energy sources is pushing technology to the nanoscale and is increasingly demanding a deeper understanding of energy transduction across all scales, particularly at the mesoscopic and nanoscale levels, where thermal fluctuations are dominant and the familiar laws of macroscopic thermodynamics are no longer sufficient. To rationally design and optimize these technologies, a new set of tools is required.

The fluctuation theorems and the broader stochastic thermodynamics toolkit provide actionable design principles for small-scale energy transduction. In practice, JE and CFT transform nonequilibrium work statistics into equilibrium free-energy differences (an indicator of the energy storage capacity of molecules and the efficiency of chemical reactions), allowing robust estimates of $\Delta F$ and $\left\langle W \right\rangle$ under sampling constraints and offering guidance to benchmark protocols \cite{ritort2008nonequilibrium,Jarzynski1997prl,crooks1998nonequilibrium,collin2005verification}. These ideas are transferable across soft-matter and mesoscopic platforms (e.g., colloids, biomolecular handles, micromachines), where finite-time protocols and dissipation matter. The same logic supports the directions in sustainable energy research, in the design of efficient artificial molecular motors, strategies to harvest ambient fluctuations, and the relationship between the meso- and macroscopic theory of chemical reaction networks \cite{Seifert2012,martinez2017colloidal,Sivak2025,Yoshimura2021}. More broadly, the framework clarifies the trade-offs between power efficiency in microscopic heat engines \cite{Benenti2017} and connects to the thermodynamics of information, where the energetic costs of measurement and computation become operationally relevant \cite{Parrondo2015,Wolpert2024}.

As the quest for sustainable energy continues to push further into the nanoscale domain, the concepts presented in this tutorial will form a valuable toolkit for scientists and engineers. Furthermore, the successful hands-on verification of fundamental fluctuation theorems, such as the Jarzynski and Crooks, highlights the relative simplicity of this approach to explore and deepen the understanding of non-equilibrium thermodynamics experimentally, hopefully helping educators to inspire new generations to advance the field further.

\vspace{2mm} 
\textbf{\textsc{Data and Code Availability:}} All the datasets used in this tutorial are openly available, with an annotated set of Python notebooks for calibrations and data analysis, on the Zenodo server at DOI:10.5281/zenodo.17201161 \cite{ZenodoDOI}.

\section{Acknowledgments}

TTM acknowledges the support from Fundação Coordenação de Aperfeiçoamento de Pessoal de Nível Superior for PROEX (Programa de Excelência Acadêmica, 88887.370240/2019-00) and CAPES-PRINT (PRINT - Programa Institucional de Internacionalização, 88887.716077/2022-00). SRM acknowledges support from FAPESP (Fundação de Amparo à Pesquisa do Estado de São Paulo), Grants no. 2019/27471-0, 2013/07276-1. A.H.A.M. acknowledges support from the National Science Centre (NCN), Poland Grant OPUS-21 (No. 2021/41/B/ST2/03207). The authors thank Leonardo Paulo Maia and Paulo Henrique Souto Ribeiro for inspiring discussions and Diogo de Oliveira Soares Pinto, Frederico Borges de Brito, and René Alfonso Nome Silva for stimulating and encouraging this line of research in São Carlos since the beginning.

\bibliography{refs}

\end{document}